\documentclass[preprint2]{aastex631}

\def\CII  {[C{\small{II}}]}
\def\micron {$\mu$m}

\shorttitle{[CII] map of NGC~7331}
\shortauthors{Sutter \& Fadda}
\graphicspath{{./}{figures/}}

\begin{document}

\title{[CII] map of the molecular ring and arms of the spiral galaxy NGC~7331\footnote{Based on observations done with FIFI-LS on SOFIA}}

\correspondingauthor{Jessica Sutter}
\email{jsutter2@usra.edu}

\author[0000-0002-9183-8102]{Jessica Sutter}
\affiliation{SOFIA Science Center, USRA, NASA Ames Research Center, M.S. N232-12 Moffett Field, CA 94035, USA}
\author[0000-0002-3698-7076]{Dario Fadda}
\affiliation{SOFIA Science Center, USRA, NASA Ames Research Center, M.S. N232-12 Moffett Field, CA 94035, USA}



\begin{abstract}
We present the [CII]~157.7~$\mu$m map of galaxy NGC~7331 obtained with FIFI-LS on SOFIA. This map extends an existent Herschel/PACS observation of the central strip of the galaxy to encompass the entire molecular ring and much of the disk, including multiple spiral arms with intense far-IR emission.
We also present Herschel archival data of the [NII]~205~$\mu$m line which covers a substantial part of the [CII] SOFIA observations and allows us to estimate the neutral fraction of the [CII] emission along the ring and disk of the galaxy. We find that the neutral fraction rises with the distance from the center. In addition, by tracing the azimuthal variation of the neutral fraction, we are able to see how our observing perspective affects this measurement. The high inclination of NGC~7331 allows us to glimpse the internal walls of the molecular ring. There, young bright stars emit UV radiation causing more [CII] emission to be produced in the ionized gas.  On the outer walls, opaque dust shrouds the rest of the ring, making the neutral medium the dominant source of [CII] emission.
Through spatial analysis comparing the [CII] emission to tracers of gas heating, we are able to investigate how the photoelectric heating efficiency varies throughout NGC~7331 and extend global measurements of the [CII] deficit to local environments. Since the origin of [CII] emission has typically been studied in face-on galaxies, our results shed a new light on the interpretation of [CII] emission especially when studying distant galaxies with unknown inclination.

\end{abstract}

\keywords{Infrared galaxies (790) --  Molecular gas (1073) -- Galaxy environments (229) -- Interstellar Medium (847)}


\section{Introduction} \label{sec:intro}
Understanding the exchange of energy in the interstellar medium (ISM) of galaxies is essential for determining how galaxies grow and evolve.  Energy injected to the ISM by young stars and the strong stellar winds they produce can spur or halt the formation of new stars through feedback processes.  Structures in galaxies like bars, rings, and spiral arms, can also influence the transportation of energy between different galactic environments.  One method for probing the detailed energetic processes within a galaxy is to compare the fluxes from a wide variety of wavelength regimes and emission lines.  Comparing the location and strength of emission from far--infrared fine structure lines that regulate cooling to indicators of dust heating or star formation, like infrared and UV emission, can provide insight into the thermal regulation of the ISM. 

Studies of this nature have been limited by the lack of galaxies for which resolved spatial coverage is available at a wide variety of wavelength regimes.  This is especially true of observations of the far-infrared (FIR) emission lines which regulate cooling processes.  Spatially--resolved observations of these lines are limited by the opacity of Earth's atmosphere to far--infrared wavelengths and the decrease in angular resolution at longer wavelengths.  Therefore, infrared observations require both space or airborne observatories with large enough detectors to resolve individual galaxies.  Currently, the main resource for achieving these observations is The Stratospheric Observatory for Infrared Astronomy \citep[SOFIA,][]{Young2012}.  As SOFIA observes from the stratosphere, it is able to observe in the infrared with minimal absorption from atmospheric water vapour.

One frequent target of extragalactic infrared studies is the 158~\micron\ line produced by singly--ionized carbon.  The \CII~158~\micron\ line is often the brightest \textit{observed} emission line in star--forming galaxies \citep[e.g.][]{Luhman1998}, making it observable in both the local and high--redshift universe.  The prominent role of \CII\ as a cooling channel for photodissociation regions (PDRs) makes it a useful indicator of thermal regulation in the ISM \citep{Wolfire2003}.  Comparing the strength of the \CII\ emission to other ISM tracers can uncover the conditions within the ISM \citep[e.g.][]{HerreraCamus2015, Cormier2019}.  

An interesting feature of the [CII]~158~\micron\ line is it multi--phase origin.  As carbon has a relatively low ionization potential of 11.3~eV (well below the 13.6~eV necessary to ionize hydrogen), singly--ionized carbon (C$^+$) is the predominate form of carbon throughout a variety of galaxy environments.  This includes both star--forming HII regions, neutral PDRs, and even some parts of molecular clouds \citep{Croxall2017,Velusamy2014}.  While the prevalence of C$^+$ contributes to the overall brightness of the [CII]~158~\micron\ line, it also complicates the use of this line as a tracer of galactic properties.  For example, as some of the [CII] emission originates in PDRs, it is often used in PDR models \citep[e.g. PDR Toolbox, see][]{Pound2008}.  For these models to accurately assess the PDR conditions, it is important that only the [CII] emission from the PDRs be considered.  This often proves problematic for the global measurements of the [CII] line, where it is typically difficult to distinguish the origin environment of the emission.  

In order to better understand \CII\ emission and how it can be used to diagnose the energy balance within the ISM, detailed observations of resolved \CII\ in local galaxies are required.  An ideal laboratory for the study of the \CII\ line is the nearby spiral galaxy NGC~7331.  Initially discovered by William Herschel in 1784, NGC~7331 has been widely studied over the past 200 years.  Located at a distance of 14.7~Mpc \citep{Freedman2001}, NGC~7331 has been included in a large number of galaxy surveys and has been mapped at wavelengths spanning X-rays \citep{Jin2019} to millimeter CO lines \citep{Leroy2009}.  Studies of NGC~7331 have found that it has a large molecular ring and $L_{\rm{FIR}}\sim2\times10^{10}L_{\odot}$ \citep{Thilker2007}.   In addition, NGC~7331 has a relatively high inclination, allowing for kinematic studies of the disk of this galaxy \citep{GarciaGomez2002}.  All of these properties make NGC~7331 an ideal target for studies of the ISM in a variety of galactic environments.  

NGC~7331 has also sparked interest as a potential Milky Way analog \citep{Thilker2007}.  The moderate star--formation rate \citep[2.74 M$_{\odot}$ yr$^{-1}$,][]{Kennicutt2011} and presence of a  bar and molecular ring make NGC~7331 a good candidate for studying what our Galaxy may look like from the outside.  In addition, with an inclination of $\sim 72$\textdegree, we see NGC~7331 from a perspective that allows us to see the interior and exterior surface of the molecular ring (see Figure~\ref{fig:HSTring} for a visualization), which can then be directly compared to the view of the Milky Way from the location of the Solar System. The presences of a star--forming molecular ring and relatively high inclination also make NGC~7331 an interesting comparison to the Andromeda galaxy \citep{Yin2009, Ford2013}.  The greater distance to NGC~7331 allows for much faster mapping of the entire disk, providing an interesting case study for better understanding the galaxies in the Local Group. 

In this paper, we present new \CII~158~\micron\ maps of NGC~7331 obtained with the Field Imaging Far-Infrared Line
Spectrometer \citep[FIFI-LS;][]{Fischer2018}.  The new FIFI-LS map expands on the existing \textit{Herschel} PACS coverage of NGC~7331.  The PACS data, part of the larger `Key Insights in Nearby Galaxies: a Far--Infrared Survey with \textit{Herschel}, or KINGFISH program \citep{Kennicutt2011}, focused on the nuclear region of NGC~7331.  As one of the primary goals of the KINGFISH survey was to understand the complex physics of star--formation, the PACS measurements from this survey centered on IR--bright nuclear and extranuclear star--forming regions.  These data provide valuable information on the local ISM conditions surrounding HII regions and PDRs \citep[see the works of][as examples]{Croxall2012, HerreraCamus2015, Smith2017, Sutter2019}.  By completing the \CII\ coverage of NGC~7331, we are able to explore the diffuse ISM surrounding the sites of active star formation targeted by KINGFISH.  This provides an important comparison as \CII\ emission can originate in such a wide variety of environments.  These spatially resolved observations from a single galaxy expand on the work done by the KINGFISH collaboration by directly comparing the properties of the \CII\ emission from different structural components and in low--luminosity regions that were not included in the KINGFISH work.  Understanding how the \CII\ emission from the diffuse ISM contributes to the integrated \CII\ emission will provide a clearer view of how the range of conditions within one galaxy effect global observations. 

By comparing the FIFI-LS \CII\ map to the wide range of available archival data, we are able to analyze the relationships between the \CII\ emission and star--formation rate indicators, dust properties, and molecular gas throughout NGC~7331.  Comparisons with available [NII]~205~\micron\ line measurements also allow us to determine what fraction of the \CII\ emission originates in ionized and neutral phases of the ISM.  Tracking the variations in the ratios of \CII\ to infrared and PAH emissions across the disk of NGC~7331 provide a case study for how galaxy structures such as rings and arms influence the photoelectric heating efficiency.  Comparisons between \CII\ and CO emission shed light on the potential conditions that could sustain CO--dark molecular gas.  This data set provides an unique opportunity to uncover the ISM conditions of a Milky Way analog and establish context for the multitude of high--redshift \CII~158~\micron\ detections.

This paper is organized as follows.  In Section~\ref{sec:data} we describe the new FIFI-LS observations and data processing as well as the archival data used to perform our analysis.  Section~\ref{sec:results} presents the comparisons between the FIFI-LS \CII\ measurements and a wide variety of galaxy properties.  In Section~\ref{sec:conclusions} we highlight the main conclusions drawn from this work.

\section{Data and Observations} \label{sec:data}
\label{sec:observations}

\begin{figure*}
\begin{center}
\includegraphics[width=0.92\textwidth]{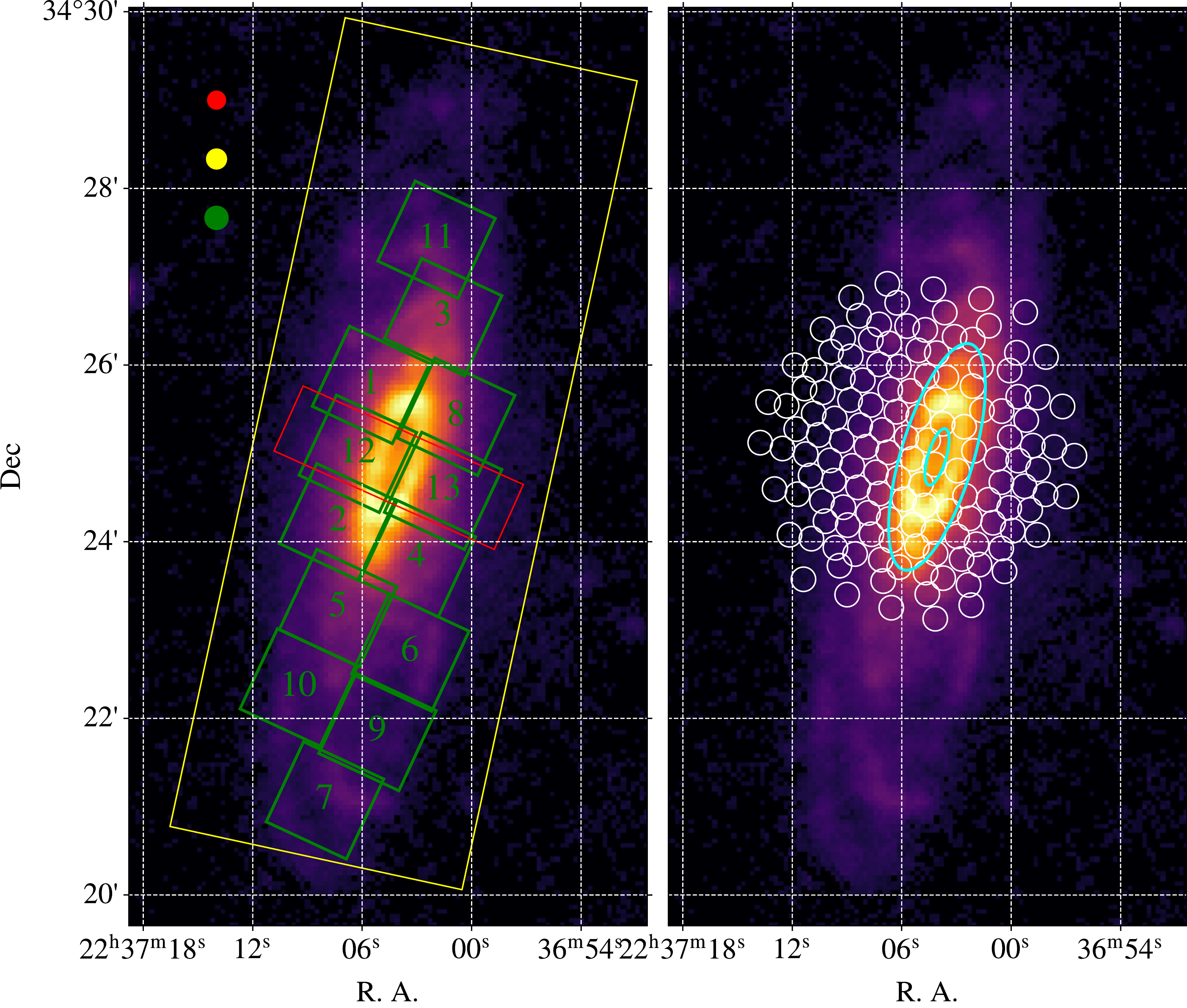}
\end{center}
\caption{Coverage of the FIFI-LS, IRAM, PACS, and SPIRE observations over the 160~$\mu$m PACS/Herschel image of NGC~7331. On the left panel, the green contours define the different FIFI-LS fields with the AOR number marked at their centers. The yellow contour shows the coverage of the IRAM CO observations and the red rectangle shows the strip covered in [\ion{C}{2}] with the PACS spectrometer. The filled circles show the beam of the FIFI-LS, [\ion{C}{2}], and CO observations with the same colors as the contours. On the right panel, the white circles show the SPIRE FTS detector beams which observed in high spectral resolution NGC~7331. The extension of the ring computed in Section~\ref{sec:lumprofile} is marked with two cyan ellipses.
}
\label{fig:coverage}
\end{figure*}
In this section we describe the new and archival data used in our analysis of NGC~7331.  This includes the new SOFIA \CII\ maps as well as Herschel spectroscopy, CO maps, and the full--wavelength range of photometry spanning FUV to far--infrared from a variety of sources.  The combination of the multiwavelength data presented here allows us to develop detailed models of the local ISM conditions across the disk of NGC~7331.  To accomplish this we use a combination of SED fits and PDR models.  SED fitting is performed using the `Code Investigating Galaxy Emission' (CIGALE) program.  As CIGALE uses an assumption of energy balance to find the best--fit SED, FUV--FIR photometric measurements are necessary.  The SED fits are then used to determine the total far--infrared emission, the amount of attenuated UV emission, and to refine measurements of the PAH emission.  Comparing these values to the \CII\ emission allow us to trace the photoelectric heating efficiency across the disk of NGC~7331.  The additional Herschel spectroscopy covers the two [NII] far--infrared fine structure lines at 122~\micron\ and 205~\micron.  The ratio of these two lines can be used to determine the electron number density in the ionized ISM.  The derived electron densities are then used to predict the ratio of \CII\ to [NII]~205~\micron\ emission, which can be used to determine the fraction of the \CII\ emission that originates in the ionized ISM.  Finally, by comparing the \CII\ and CO emission, we can determine if there is a significant reservoir of CO dark gas in NGC~7331 and determine the conditions within the PDRs where a large fraction of the \CII\ emission originates.  The following subsections describe these data sets in further detail and provide an account of the processing steps performed to extract the line and photometric fluxes.

\subsection{SOFIA/FIFI-LS spectra}
\label{sec:FIFI-LS}

\begin{deluxetable*}{cccccccccc}[!t]
\setlength{\tabcolsep}{2pt}
\tablecolumns{10}
\tablewidth{0pt}
\tablecaption{ Log of FIFI-LS observations \label{tab:log}}
\tablehead{
\colhead{Observation}\vspace{-0.2cm} &\colhead{Flight}
& \colhead{AOR} & \colhead{R.A.} & \colhead{Dec} & \colhead{Starting} &
\colhead{Exposure}&\colhead{Barometric} &\colhead{Zenithal} & \colhead{Zenithal} \\
\colhead{Date}\vspace{-0.2cm}&\colhead{Number}
& \colhead{ID}  & \multicolumn{2}{c}{[J2000]} &\colhead{Time} &
\colhead{Time}& \colhead{Altitude} & \colhead{Angle} & \colhead{Water vapor}\\ 
\colhead{[UT]} &  
&  & \multicolumn{2}{c}{[degs]}& \colhead{[UT]} &
\colhead{[minutes]} &\colhead{[ft]} & \colhead{[degs]} & \colhead{[$\mu$m]}
}
\startdata
2019 10 31 & 632 & 75\_0045\_1 & 22.6182&34.4296 &07:09:22 & 21 & 41000 & 32.1 -- 36.7 & 4.0 -- 4.4\\
2019 10 31 & 632 & 75\_0045\_8 & 22.6169&34.4236 &07:33:10 & 21 & 41000 & 37.5 -- 41.4 & 4.6 -- 4.9\\
2019 10 31 & 632 & 75\_0045\_2 & 22.6187&34.4037 &07:55:00 & 21 & 41000 & 42.6 -- 45.9 & 4.7 -- 4.9\\
2019 10 31 & 632 & 75\_0045\_4 & 22.6175&34.3969 &08:16:25 &  9 & 41000 & 46.1 -- 47.5 & 4.7\\
2019 11 02 & 634 & 75\_0045\_4 & 22.6175&34.3969 &07:22:09 & 12 & 43000 & 36.3 -- 39.1 & 4.3 -- 5.0\\
2019 11 02 & 634 & 75\_0045\_3 & 22.6171&34.4424 &07:37:04 & 15 & 44000 & 39.8 -- 43.1 & 3.2 -- 3.6\\
2019 11 14 & 640 & 75\_0045\_12& 22.6184&34.4166 &05:31:57 & 21 & 39000 & 33.6 -- 37.5 & 7.7 -- 7.8\\
2019 11 14 & 640 & 75\_0045\_13& 22.6171&34.4096 &05:53:35 & 21 & 39000 & 37.8 -- 41.7 & 7.5 -- 7.7\\
2019 11 14 & 640 & 75\_0045\_5 & 22.6187&34.3875 &06:14:45 & 21 & 39000 & 41.7 -- 45.7 & 7.6 -- 8.0\\
2019 11 14 & 640 & 75\_0045\_6 & 22.6176&34.3791 &06:37:25 & 23 & 39000 & 45.7 -- 49.8 & 5.8 -- 7.8\\
2019 11 14 & 640 & 75\_0045\_9 & 22.6181&34.3641 &07:01:29 & 10 & 39000 & 49.9 -- 51.6 & 5.4 -- 5.7\\
2020 09 03 & 681 & 75\_0045\_3 & 22.6172&34.4434 &09:32:03 &  6 & 43000 & 36.0 -- 36.8 & 6.4 -- 6.8\\
2020 09 03 & 681 & 75\_0045\_11& 22.6172&34.4570 &09:38:33 & 20 & 43000 & 36.9 -- 41.1 & 5.5 -- 6.8\\
2020 09 03 & 681 & 75\_0045\_10& 22.6193&34.3725 &09:59:59 & 20 & 43000 & 41.6 -- 45.5 & 4.2 -- 5.3\\
2020 09 03 & 681 & 75\_0045\_9 & 22.6181&34.3641 &10:20:53 & 21 & 43000 & 45.9 -- 50.1 & 4.6 -- 5.3\\
2020 09 03 & 681 & 75\_0045\_7 & 22.6189&34.3512 &10:42:13 &  4 & 43000 & 50.2 -- 50.9 & 5.0 
\enddata
\tablecomments{Observations are presented in chronological order. Observations on the same date were made during a single flight leg. The zenithal angle varied linearly during the observations between the two reported values.}
\end{deluxetable*}

The SOFIA observations were performed as part of the discretionary directorial time (PLAN ID 75\_0045, P.I. Fadda) awarded after a flash call to fill a particular direction in the sky during SOFIA Cycle~7. The Field Imaging Far-Infrared Line Spectrometer \citep[FIFI-LS,][]{Fischer2018, 2018JAI.....740004C} was used to map the [\ion{C}{2}]~157.741~$\mu$m (rest frame) line. FIFI-LS is an infrared integral field spectrometer with two channels capable of covering a wavelength range from 50 to 125~$\mu$m (blue channel) and 105–200~$\mu$m (red channel). It has an array of 5$\times$5 spatial pixels covering a field of view of 0.5$\times$0.5 sq arcmin. in the blue (6$\times$6 sq arcsec per spatial pixel) and 1x1~sq arcmin (12$\times$12~sq arcsec per spatial pixel) in the red. For NGC~7331, the [\ion{C}{2}]~157.741~$\mu$m line corresponds to the observer frame wavelength of 158.101~$\mu$m. At this wavelength, the spectral resolving power of FIFI-LS is 1155, meaning that an unresolved line has a FWHM of 260~km~s$^{-1}$. The spatial resolution of the instrument is 15.6~arcseconds, corresponding to 1.1~kpc at the distance of the NGC~7331 galaxy \citep[14.7~Mpc,][]{Freedman2001}. FIFI-LS is a dual channel instrument. Parallel observations were obtained at 88.3~$\mu$m. Unfortunately, the 88.3~$\mu$m coverage is incomplete since the blue channel was malfunctioning during the last flight. Since the depth is insufficient to derive any science results, the blue channel data are not discussed in this paper.

The data were acquired in four flights (2019 Oct 31, 2019 Nov 2 and 14, 2020 Sep 3) for a total of approximately 4.5~hr of flight time (see Table~\ref{tab:log}). The map was divided in 13  Astronomical Observation Requests (AORs) to cover NGC~7331's ring, disk, and arms. The coverage is uniform over the entire field with the exception of the last field observed (\# 7 in Figure~\ref{fig:coverage}) whose observation was not completed.

Figure~\ref{fig:coverage} shows the locations of the 13 AORs superimposed on the PACS~160~\micron\ image.  Additionally, the spatial coverage of the IRAM CO observations (yellow region) and \textit{Herschel} PACS observations (red region) are displayed for reference. The FIFI-LS observations were performed in the chop--nod mode with the secondary mirror chopping between the galaxy and reference fields on the two sides of the galaxy, each at a 270~arcsecond distance from the major axis of the galaxy. 

Some dithering was performed to reduce the effect of bad pixels and to improve the recovery of the point spread function (PSF) in the images, since the size of the spatial pixel of FIFI-LS (12\arcsec) is not small enough to recover the shape of the PSF. The data were reduced using the FIFI-LS pipeline \citep{2020ASPC..Vacca}.  We computed the errors from the dispersion of the measurements in each pixel of the map, rather than using the errors propagated by the pipeline which are based on formal errors of the slopes of the signal ramps. Finally, the data were corrected for atmospheric transmission using the ATRAN model \citep{1992nstc.rept.....L}. The values of the zenithal water vapor burden were estimated through satellite observations \citep{iserlohe2021} and rescaled to values measured during the flights.  Because of the good atmospheric transmission in the wavelength range considered (95\% on average), this correction contributes in a negligible way to the error of the final flux.

The reduced data were projected into spectral cubes with a fine grid of 3\arcsec\ sampling using a Gaussian spectral kernel with a dispersion equal to 1/4 of the spectral resolution  and a Gaussian spatial kernel with a dispersion equal to 1/2 the spatial resolution. These parameters produced a data cube that conserves the instrumental spectral and spatial resolutions.  Examples of spectra extracted from different apertures can be found in the Appendix~\ref{sec:PACSvsFIFI}, which presents a comparison between FIFI-LS and PACS fluxes.

\subsection{Herschel/PACS spectra}
\label{sec:PACS}
A 40"$\times$2'40" strip across the center of NGC~7331 oriented perpendicularly to the major axis of the galaxy was covered with the PACS spectrometer on board Herschel (see Figure~\ref{fig:coverage}).
Several lines were observed as part of the KINGFISH Herschel key project, including [CII]~157.7~$\mu$m (AOR ID 1342222579) and [NII]~122~$\mu$m (AOR ID 1342222578) used in our analysis. The observations were performed in {\it unchopped spectroscopy} due to the required nodding distance being beyond the limits of the PACS spectrometer. Such observations are known to be very sensitive to transients, variations of the response of the detectors due to sudden change of charges on them~\citep{Fadda2016} which can be caused by sudden variations of flux or by cosmic ray events.

The pipeline products available in the Herschel Science Archive (HSA) are not corrected for transients. Moreover, the HSA products are calibrated only using the calibration blocks of the observation. Since the calibration is performed at the beginning of the observations by alternately observing two internal black bodies at different temperatures, this produces transients during the calibration and along the entire observation. For this reason, all the chopped observations in the HSA have been calibrated using the telescope background as a reference. This has been accurately characterized during the mission and shown to be very stable.
Unfortunately, the same technique has not been applied to the unchopped observations.

We reduced the data using the transient correction pipeline available in HIPE~15 \citep{Fadda2016} which corrects for transient behaviours and calibrates the flux using the telescope background as reference. We cross-correlated our reduction with the fluxes published in \citet{Croxall2017} and in the case  of the [NII]~122~$\mu$m observations finding a difference of approximately 20\% in the fluxes. We refer to the appendix (section~\ref{sec:NIIdata}) for a discussion about the problems of this kind of data and the possible origin of the difference.

In order to assess the calibration of the [CII] data, the central part of the galaxy was re-observed with SOFIA. This allowed us to cross-correlate the flux calibration of SOFIA against Herschel. As more extensively explained in the appendix (see Section~\ref{sec:PACSvsFIFI}), the Herschel data taken directly from the archive are not calibrated using the telescope background as absolute calibrator. This leads to a difference in fluxes of about 30\%, which could be related to the 30\% uncertainty described in \citet{Croxall2012}. Once this calibration is taken into account using the transient correction pipeline in HIPE~15, the fluxes agree very well with the SOFIA measurements.

Although PACS observations at 205~$\mu$m exist, they are strongly affected by a filter leakage. For this reason, we consider in our analysis the SPIRE [NII]~205~\micron\ data which have the additional advantage of covering a larger region of the galaxy.

\subsection{Herschel/SPIRE spectra}
NGC~7331 has been observed with the Fourier Transform Spectrometer (FTS) of SPIRE  in intermediate point mode and high spectral resolution (AOR ID 1342245871). This observation mode consists of four jiggling positions of the detectors which allow an uniform coverage of an hexagonal area (see Fig.~\ref{fig:coverage}, right panel). Since two detectors were damaged, a few holes are left in the coverage map.  From this data set, we use the [NII]~205~$\mu$m line which is well detected across NGC~7331, especially along the dusty ring. The intensity of this line is used to estimate the fraction of the observed [CII] emission coming from the ionized and neutral medium.

The archival data have been reprocessed through HIPE 15.0.1 using the latest available SPIRE calibration (version 14.3) and the flux calibration for extended emission. It is important to notice that this version of the SPIRE calibration corrects for an important error in extended source flux calibration discovered by \citet{Valtchanov2018}. This error was corrected with version 14.2 of the SPIRE calibration in Nov 2015 and used in the developer version of HIPE~14~\footnote{\url{herschel.esac.esa.int/hcss-doc-15.0/load/spire_drm/ spire/release/calibration/spire_cal_14_2.html}}. Results published before this date usually underestimate the total flux of SPIRE lines by a factor of~1.3--2. In particular, the release note of HIPE~14.1~\footnote{
\url{www.cosmos.esa.int/documents/12133/1561728/Old+What\%27s+new+in+version+14.x}} explains that the correction leads to systematic $\approx$40\% higher intensities for SSW, the channel used to measure the [NII]~205~$\mu$m line.

Due to the difference in sizes of SOFIA and Herschel telescopes, the SPIRE beam at 205~$\mu$m approximately corresponds to the beam of FIFI-LS at 158~$\mu$m \citep[16.6~arcsec according to][]{Makiwa2013}. Since the data have been taken in high resolution mode, at 205~$\mu$m the spectral resolution is 297~km/s.

\subsection{Infrared imaging}
The galaxy has been observed with the IRAC \citep{2004ApJS..154...10F} and MIPS \citep{2004ApJS..154...25R} instruments aboard the {\it Spitzer Space Telescope} \citep{2004ApJS..154....1W}. We retrieved the relevant data from the {\it Spitzer} Heritage Archive. The quality of the archival data was sufficient for our analysis.

NGC~7331 was observed by {\it Herschel} with the PACS and SPIRE instruments as part of the KINGFISH program \citep{Kennicutt2011}. In this paper we will consider only the PACS observations obtained at 70, 100, and 160~$\mu$m and the 250~$\mu$m image obtained with SPIRE. We do not consider the other SPIRE images, since their spatial resolution is too low in comparison to our SOFIA data. Moreover, the emission at those wavelengths is well beyond the peak of the infrared emission and not useful to constrain the total infrared emission. We did not reprocess the images for our work since the quality of the archived products in the {\it Herschel} Science Archive is adequate for our analysis.  The fluxes determined from these data are used for SED modeling to determine further properties of NGC~7331.

Also the Infrared Space Observatory ({\sl ISO}) observed a strip covering the whole ring of NGC~7331 with the ISOCAM instrument in several bands~\citep{BJSmith1998}. In this paper, we make use of the LW8 band observation to probe the strength of the 11.3~$\mu$m PAH feature. For our analysis, we used the calibrated image from the {\sl ISO} ESA archive. 
\begin{figure*}[!t]
\begin{center}
\includegraphics[width=\textwidth]{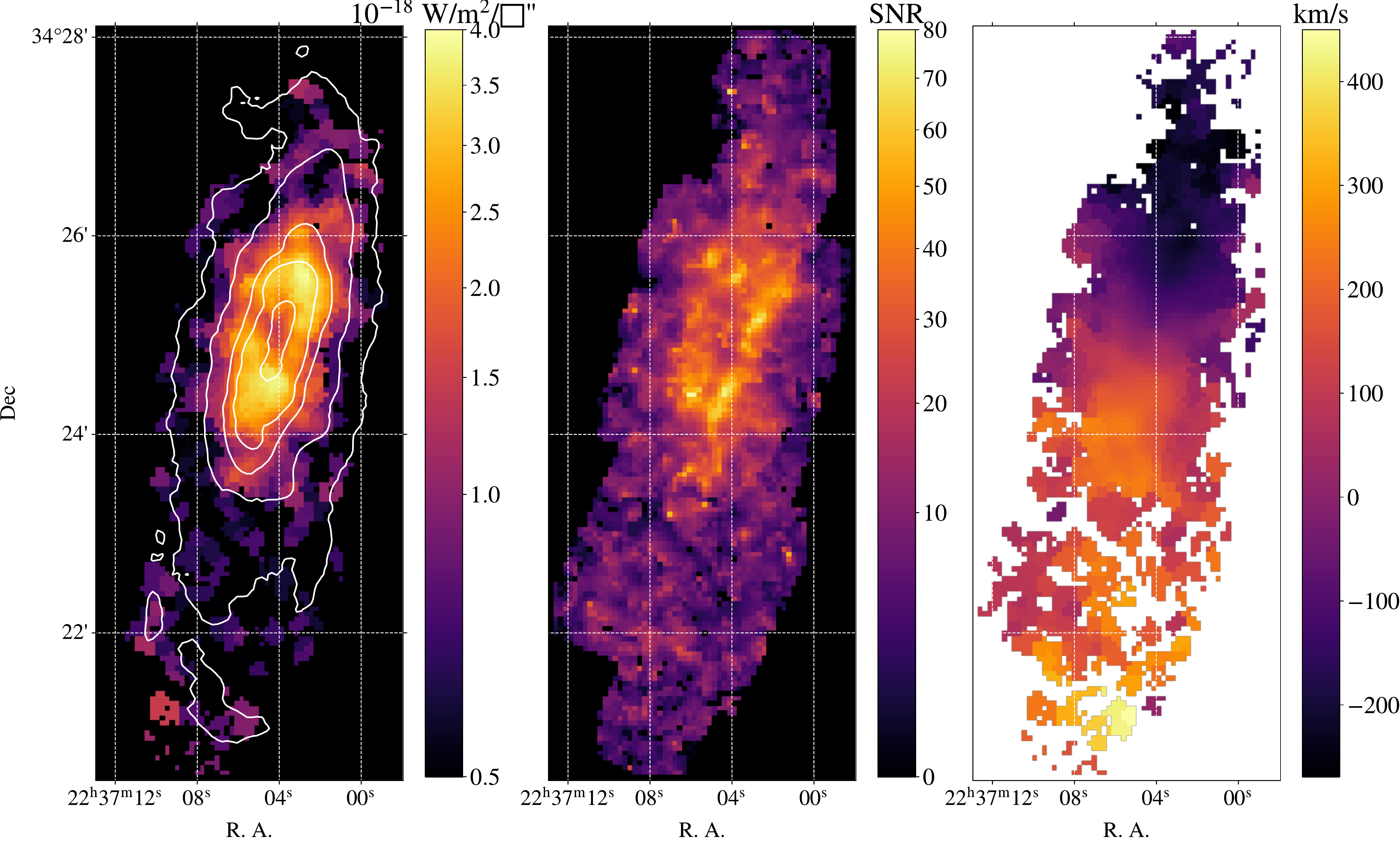}
\end{center}
\caption{[CII] map of NGC~7331. On the left, contours (0.4, 1.3, 3.4, 6.4~MJy/sr) of the PACS 160~$\mu$m image overlapped on the [CII] map to show the location of ring and arms. The central map show the signal-to-noise for the line fits of each pixel of the spectral cube.
On the right, the velocity map assuming a systemic velocity of 800~km/s. Pixels with signal-to-noise ratio lower than 5 are left blank.
}
\label{fig:CIImap}
\end{figure*}

\subsection{CO data}
\label{sec:COdata}
Observations of the $^{12}$CO$_{2\rightarrow 1}$ line at 230~GHz were obtained with the 30m IRAM telescope in Granada, Spain, as part of the `The HERA CO-Line Extragalactic Survey’ (HERACLES)~\citep{Leroy2009} in October 2007 and January 2008. Reduced data were downloaded from the survey website~\footnote{\url{www.iram.fr/ILPA/LP001}}. The size of the IRAM beam is 13.4~arcsec. The spectral resolution of the observation is 2.6~km/s.
Data are expressed in main beam temperatures and binned in channels of 10.4~km/s.
To convert main beam temperatures ($T_{MB}$) into fluxes, we assume that the extended source is constant over the beam and used the formula in the IRAM spectral line calibration report \citep{Kramer1997}:
\begin{equation}
    S_{\nu} [Jy] = \frac{2 k}{A} \frac{F_{eff}}{\eta_A} T_A^*[K] = 3.906 \frac{B_{eff}}{\eta_A} T_{MB} [K]
\end{equation}
We assumed the main beam efficiency $B_{eff} = 0.58$ reported in the header of the FITS file containing the data and the aperture efficiency $\eta_A = 0.41$ obtained by using the Ruze formula with values for IRAM~30m after March 2005~\footnote{\url{web-archives.iram.fr/IRAMFR/ARN/aug05/node6.html}}:
\begin{equation}
    \eta_A = 0.63 \, e^{-(\frac{4\pi\, 67.4}{\lambda[\mu m]})^2}
\end{equation}

Observations of the $^{12}$CO$_{1\rightarrow 0}$ line at 115~GHz were obtained with the BIMA array and the NRAO 12m telescope \citep{Helfer2003}. 
The major and minor axes of the beam are 6.1 and 4.9 arcsec.
The spectral resolution is 10~km/s.
Data were downloaded from the survey page~\footnote{\url{ned.ipac.caltech.edu/level5/March02/SONG/SONG.html}}. The cube units are Jy/beam and the velocity bins correspond to the spectral resolution.

In our analysis we use only the HERACLES observations due to their greater depth compared to BIMASONG observations. Since we compare the $^{12}$CO$_{1\rightarrow 0}$ to the [CII] emission, a conversion factor between the line strength of the two rotational levels $^{12}$CO$_{2\rightarrow 1}$ and $^{12}$CO$_{1\rightarrow 0}$ is required.  We determine this conversion factor through comparisons of the HERACLES and BIMASONG observations.
To compare the two observations, we converted the BIMASONG
fluxes into brightness temperatures using the Rayleigh-Jeans equation:

\begin{equation}
    T [K]= \frac{\lambda^2}{2 k \Omega} S = \frac{2\ln 2}{\pi k} \frac{\lambda_{rest}^2}{\theta_m \theta_M} I = 3.075 \, I [\frac{Jy}{beam}]
\end{equation}
with $k$ the Boltzmann constant and $\theta_m$, $\theta_M$, the axes of the BIMASONG beam.
After smoothing the two observations to the same spatial and velocity resolution, we compared the peak temperatures and the integrated temperatures of the CO lines across the galaxy. The median ratio of peak temperatures is 0.57, while the median ratio of the integrated temperatures in velocity along the line is 0.61.  The region where the comparison is possible is restricted to the ring of the galaxy. Across this region, the ratio is constant except for the nuclear region where it is slightly higher. Beyond the ring, the BIMASONG data are not deep enough to allow a direct comparison with the HERACLES data.
These values agree very well with those reported in literature~\citep{Leroy2009}. If one integrates in frequency, the typical ratio between $^{12}$CO$_{2\rightarrow 1}$ and $^{12}$CO$_{1\rightarrow 0}$ is 1.2, since $\frac{d\nu}{dv} = \frac{\nu_{rest}}{c}$ and the ratio of the frequencies of the two lines is 2. 

\subsection{UV, visible, Near-IR Data}

Spectral energy distribution models (SEDs) were created for individual regions in NGC~7331 using the wealth of available UV through IR photometry. These observations include the two {\it GALEX} bands (FUV and NUV), SDSS images in the five Sloan bands ($u'$, $g'$, $r'$, $i'$, $z'$), and 2MASS images in the $J$, $H$, and $K_{\rm s}$ bands. The {\it GALEX} FUV observation are from the Nearby Galaxy Atlas survey~\citep{GildePaz2007}. Images and spectra were retrieved from the respective archives: MAST, SDSS, and IRSA.
Foreground stars and background galaxies were removed from the SDSS images using the process described in \citet{Cook2014}. All data were smoothed to a uniform resolution of 15.6\arcsec\ to match the FIFI-LS observations.  


\section{Results and Discussion}
\label{sec:results}
With the full suite of archival data and the completed \CII\ map of NGC~7331, explorations of the ISM conditions in different environments can be performed.  In order to determine how the \CII\ emission varies across the ring, disk, and nucleus we first explain how we distinguish these locations.  To build a picture of what properties could shape observed variations across the disk, we next model the SEDs to determine important factors that could limit or increase the brightness of the \CII\ line.  Using the information provided through the SED fits, we are then able to trace how indicators of heating relate to \CII, which is widely regarded as one of the primary cooling channels of PDRs. In addition we use the information from the available [NII] 122~\micron\ and 205~\micron\ line detections to model the conditions in the ionized ISM and determine what fraction of the \CII\ emission  originates co--spatially with the [NII] emission in ionized phases of the ISM. Finally, by comparing the \CII\ emission to the CO emission, we are able to search for CO--dark molecular gas and determine the properties of the PDRs that produce much of the \CII\ emission.  Throughout these analyses, we compare to measurements of star--forming regions from the KINGFISH survey and global measurements from surveys like the Great Observatories All-sky LIRG Survey, where data is available.  The inclusion of the new NGC~7331 data expands the range of conditions in which the production of the \CII\ line has been examined and helps establish how detections of this line might be used in future observing campaigns.

\subsection{[CII] map}
\label{sec:CIImap}

The map of the integrated [CII] emission is presented in the left panel of Figure~\ref{fig:CIImap}.  This map was created by fitting the [CII] line in each pixel of the spectral cube. The contours of the PACS 160~$\mu$m image are overlapped to show the structure of the galaxy. The ring, some structure of the disk, and a few spots of the arms have strong [CII] emission. The velocity map, in the right panel, clearly displays the rotation of the galaxy. In the southern part of the spectral cube, the [CII] lines are weaker and the signal-to-noise is sometimes lower than 5. These pixels are left blank in the map.
In the rest of the analysis, we will consider fluxes of the [CII] line extracted in circular apertures across the different environment of the galaxy, each with diameter equal to the FIFI-LS beam.


\begin{figure}[!t]
\begin{center}
\includegraphics[width=0.49\textwidth]{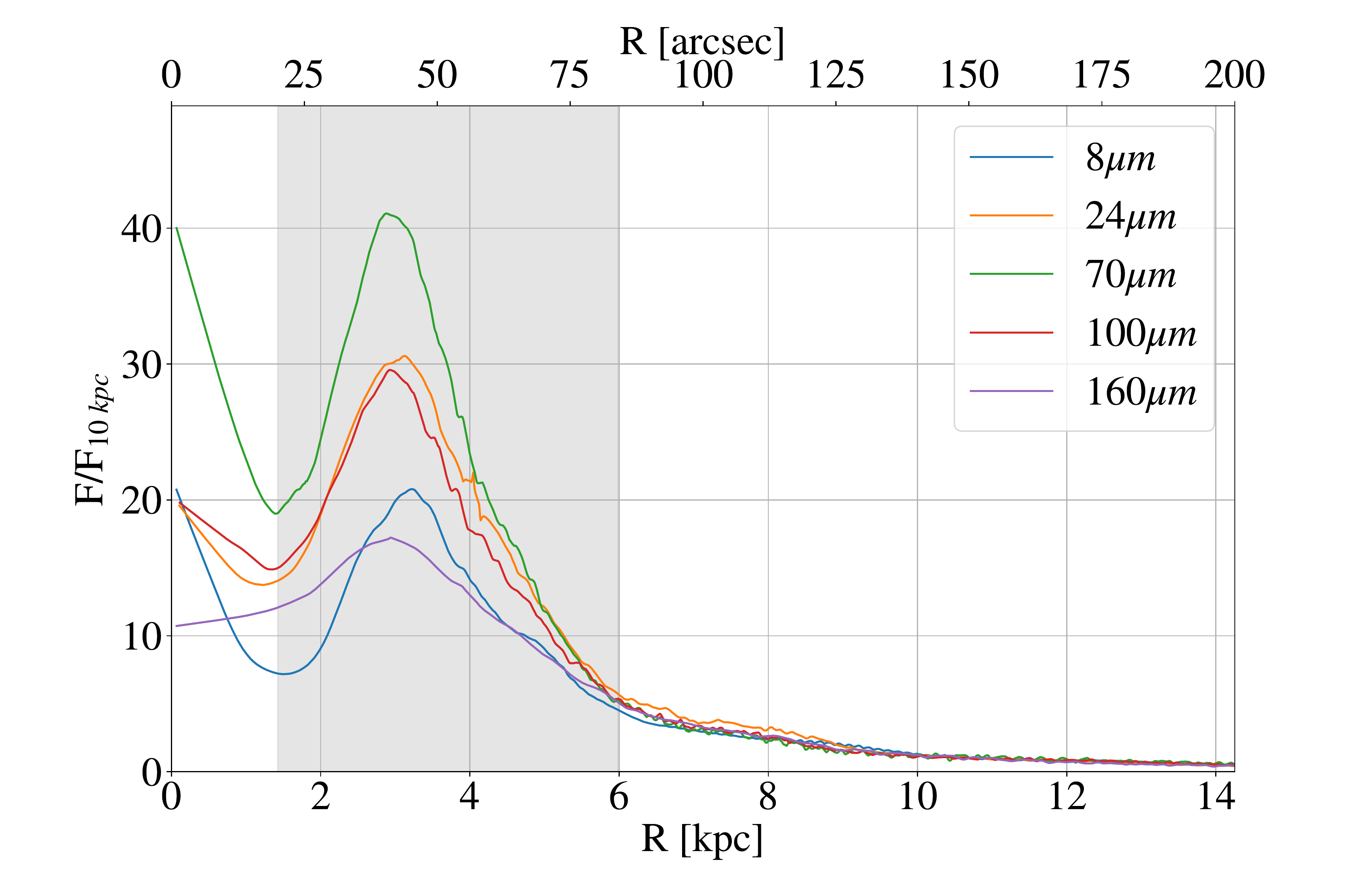}
\end{center}
\caption{
Luminosity profile as function of the deprojected radius in different mid- to far-IR channels. Each curve has been normalized to the value at 10~kpc. The bump due to the presence of the dusty ring is clearly visible in all the infrared channels in a region between 20 and 80~arcsec from the galaxy center.
}
\label{fig:lumprofile}
\end{figure}

\subsection{Luminosity profile and location of the ring}
\label{sec:lumprofile}
In order to compare the [CII] emission from isolated environments across NGC~7331, we must first determine how to identify regions of interest across our data.  Based on the photometric data, we choose to divide regions into five main categories: those interior to the molecular ring (i.e. the core), those within the molecular ring, those within the flocculant disk surrounding the ring, those within a spiral arm, and those in the intra--arm disk.  To determine the placement of each region, we must first specify how we define the molecular ring.  In this section we will explain our method for defining the ring by first determining the luminosity profile of NGC~7331 as a function of deprojected distance from its center and using this profile to identify the location of the molecular ring. 

To compute the deprojected distance from the center of the galaxy we first assume the center measured at the position of the nucleus of the galaxy at 8~$\mu$m: $22^h37^m04.06^s+34^o24'56.89''$~(J2000). If we call $\phi_0$ the position angle of the galaxy and $i$ its inclination, a point at projected distance $\rho$ from the center and projected angle $\phi$ will have a deprojected distance equal to:
\begin{equation}
R = \frac{\rho}{cos(i)} \sqrt{ \sin^2(\phi - \phi_0) + \cos^2(i) \cos^2(\phi - \phi_0)},
\end{equation}
and a deprojected angle with respect to the major axis:
\begin{equation}
\Phi = \arctan \frac{\tan(\phi-\phi_0)}{\cos i}.
\end{equation}

Using these formulae we can plot the flux of each pixel of the infrared images as a function of the deprojected radius. Position angle $\phi_0$ and inclination $i$ can be found by minimizing the dispersion of the points in the plots. We obtained the minimum dispersion for $\phi_0 = 73.5^o$ (measured West to North) and $i = 72^o$ which are consistent with values reported in literature \citep[e.g.][]{Ma1998}.
The luminosity profiles of several mid- to far-IR channels are shown in Figure~\ref{fig:lumprofile} normalized at the radius of 10~kpc. A clear bump in the luminosity profile occurs between 20\arcsec\ and 84\arcsec\ (shaded region in figure), which roughly corresponds to distances of 1.5 and 6~kpc at the distance of NGC~7331. We adopted these two values as internal and external radii of the molecular dusty ring.
The ellipses corresponding to these radii are traced in Figure~\ref{fig:coverage} over the 160~$\mu$m emission.
\begin{figure*}[!t]
\begin{center}
\includegraphics[width=0.9\textwidth]{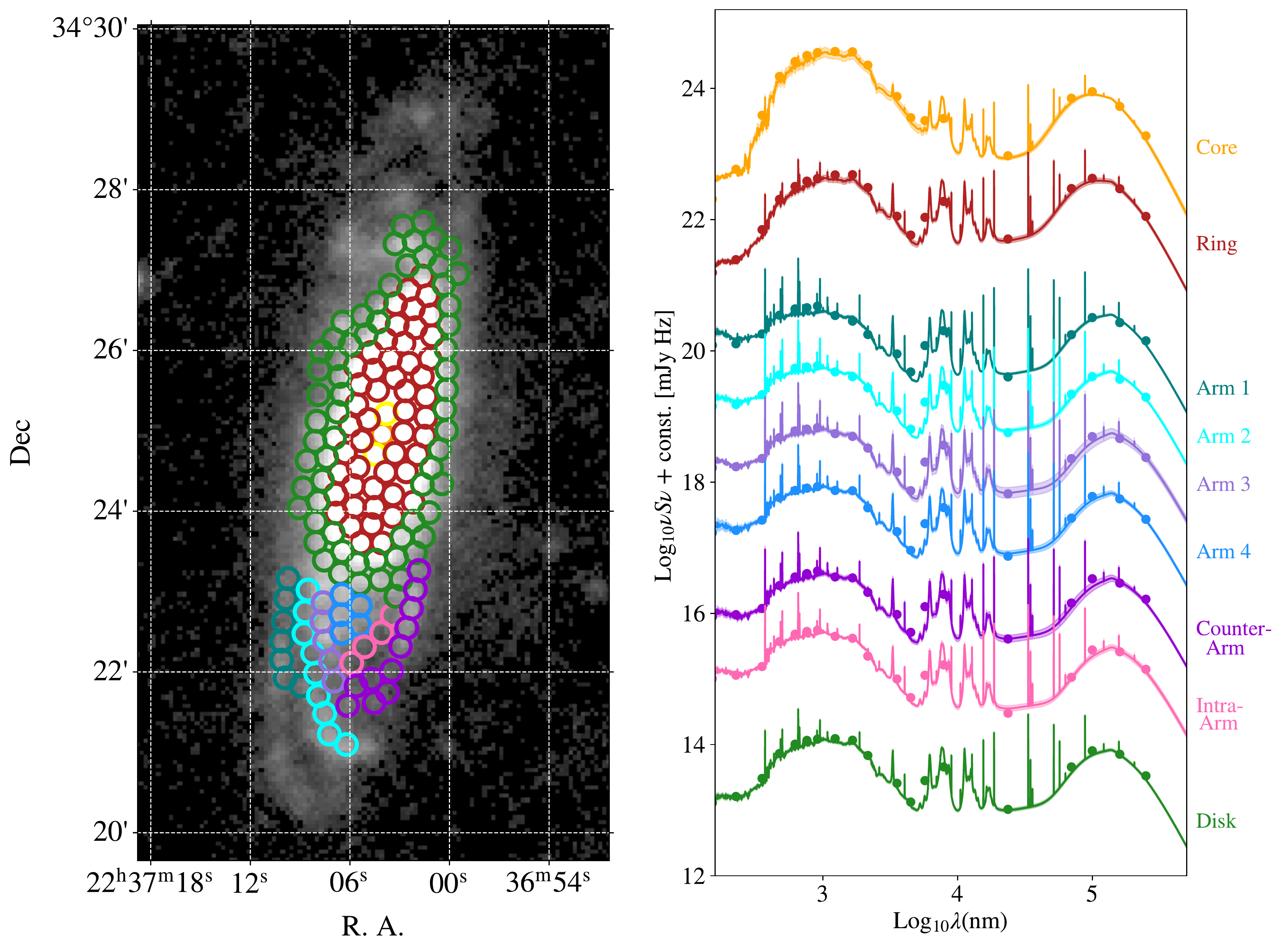}
\end{center}
\caption{\textit{Left:} The IRAC 8.0~\micron\ map of NGC~7331 with the apertures used throughout this work. The different colors corresponds to different environments within NGC~7331. Each aperture has a diameter of 15.6\arcsec\ to match the FIFI-LS spatial resolution. \textit{Right:} Average SED models and photometric data used to determine the models for the different environments color coded as in the left panel. The SEDs have been separated to better appreciate differences in shape across different environments. The standard deviation for each individual components is shown as a shaded band over each fit.}
\label{fig:SEDfits}
\end{figure*}

\subsection{Distinguishing different environments}
\label{sec:regions}
With the determination of the location of the molecular ring and the orientation of NGC~7331, we can begin to identify ideal locations for region placement across the disk.  These regions are used throughout the text as apertures for flux extraction except when we use the [NII]~205~\micron\ data where the SPIRE detectors are used instead as the SPIRE--FTS detections do not span the full disk.  All regions have a diameter of 15.6\arcsec\ to match the spatial resolution of the FIFI-LS observations.  

The chosen apertures are shown in the left panel of Figure~\ref{fig:SEDfits}.  We set three regions interior to the molecular ring, one centered on the nucleus (orange region) and one each on either side of the nucleus along the major axis (yellow regions).  Outside the nucleus, 51 red regions fill the molecular ring, defined using the method described in Section~\ref{sec:lumprofile}.  The rest of the area mapped by FIFI-LS is divided into five spiral arms, the intra--arm medium, and the flocculant disk outside of the ring where there were no clearly defined spiral arms.  Visually, we select regions to follow five spiral arms.  Four of the arms appear to follow the rotation of the disk and are labeled `Arms 1--4' in Figure~\ref{fig:SEDfits}.  One additional arm--like structure that does not seem to trace the overall disk rotation is identified.  This arms is labeled `Counter Arm,' based on its apparent reverse direction with respect to the four other arms.  Finally, four pink regions fill an area between the arms we classify as the `Intra--Arm' material.  These distinctions allow us to determine how the arms, disk, ring, and nucleus of NGC~7331 display different environmental properties.

\subsection{SED Modeling}
\label{sec:SED}
\begin{deluxetable}{lccl}
\tablecaption{Bands used for SED Fitting}
\tablehead{\colhead{Filter} &\colhead{Wavelength} &\colhead{$\sigma_{\rm{cal}}$} & \colhead{Ref$^a$} \\ 
\colhead{} & \colhead{ $\mu$m } & \colhead{} & \colhead{}} 
\startdata
GALEX\_FUV & 0.152 & 0.05 mag & 1\\
GALEX\_NUV & 0.227 & 0.03 mag & 1 \\
SDSS\_$u$ & 0.354 & 2\% & 2 \\
SDSS\_$g$ & 0.477 & 2\% & 2 \\
SDSS\_$r$ & 0.632 & 2\% & 2 \\
SDSS\_$i$ & 0.762 & 2\% & 2 \\
SDSS\_$z$ & 0.913 & 2\% & 2 \\
2MASS\_$J$ & 1.235 & 0.03 mag & 3 \\
2MASS\_$H$ & 1.662 & 0.03 mag & 3 \\
2MASS\_$Ks$ & 2.159 & 0.03 mag & 3 \\
IRAC\_CH1 & 3.550 & 1.8\% & 4 \\
IRAC\_CH2 & 4.490 & 1.9\% & 4 \\
IRAC\_CH3 & 5.730 & 2.0\% & 4 \\
IRAC\_CH4 & 7.870 & 2.1\% & 4 \\
MIPS\_24 & 23.70  & 4.0\% & 5\\
PACS\_70 & 71.11 & 5\% & 6 \\
PACS\_100 & 101.20 & 5\% & 6 \\
PACS\_160 & 162.70 & 5\% & 6 \\
SPIRE\_250 & 249.40 & 4\% & 7 \\
\enddata
\label{tab:sedbands}
\tablecomments{$^{(a)}$ References: (1) \citet{Morrissey2007}, (2) \citet{Padmanabhan2008}, (3) \citet{Skrutskie2006}, (4) \citet{Reach2005}, (5) \citet{Engelbracht2007}, (6) \citet{Balog2013}, (7) Spire Observers' Manual}
\end{deluxetable}

Once all photometric data had been smoothed to the same spatial resolution and processed, Spectral Energy Distribution (SED) fitting was performed using the Code Investigating GALaxy Evolution \citep[CIGALE][]{Noll2009, Boquien2019}.  CIGALE models the SED by assuming the energy absorbed by dust from the UV to the near--infrared is balanced by the energy emitted by dust in the mid and far--infrared.  The full suite of photometric observations used to determine the SED fits are described in Section~\ref{sec:observations} and are listed in Table~\ref{tab:sedbands} along with the calibration uncertainties used to estimate the errors.   Fits were determined using the \citet{BruzualCharlot2003} stellar population and the \citet{Draine2014} dust models.  A complete list of the parameters and modules used in the SED models can be found in Table~\ref{tab:sedprops}.  SED fits were performed on the apertures discussed in Section~\ref{sec:regions}. Errors were determined using the sum in quadrature of the variation of the sky brightness and the calibration uncertainty for each photometric detector listed in Table~\ref{tab:sedbands}. These SED models allow for estimations of the dust properties and attenuation rates in different environments in NGC~7331.  Average SED models for the nucleus, the molecular ring, the flocculant disk, the individual arms, and the intra--arm material are shown in the right panel of Figure~\ref{fig:SEDfits}.

The variation of the SEDs across the different environments of NGC~7331 can be further explored by examining the components of the CIGALE fits. The core is dominated by light from old stars, representing 78\% of the total luminosity, while the ring, disk, and arms all show less than 50\% of the total luminosity from this component.  The luminosity in the ring is primarily from dust emission, with approximately 52\% of the total luminosity coming from dust.  The disk also shows a relatively large dust luminosity component at 42\%, compared to the arms and core which have 23\% and 20\%, respectively.  The spiral arms show the highest fraction of emission from young stars, at 8\% of the total luminosity, while the ring and core have almost no observed emission from young stars (2\% for the ring and less than 1\% for the core).  These differences are visible in the changing shape of the UV emission indicating younger stars and the height and location of the infrared bump indicating changes in the dust emission.  Taking all of this together we can see how varied the environments are across NGC~7331.  The results of the SED fits are used to derive the indicators of heating used in the following subsection as comparisons to the cooling done by \CII.


\begin{deluxetable}{ll}
\tablecaption{Parameter values for CIGALE modules}
\label{tab:sedprops}
\renewcommand{\arraystretch}{0.87}
\tablehead{\colhead{Parameter} &\colhead{Input values} } 
\startdata
\multicolumn{2}{c}{\texttt{sfhdelayed}}\\
\texttt{tau\_main} [Gyr] & [0.5, 10], $\delta=0.25 $ \\
\texttt{age\_main} [Gyr] & 11 \\
\texttt{tau\_burst} [Gyr] & 0.05 \\
\texttt{age\_burst} [Gyr] & 0.02 \\
\texttt{f\_burst} & 0.0 \\
\texttt{SFR\_A} [M$_{\odot}$/yr] & 1.0 \\
\hline
\multicolumn{2}{c}{\texttt{bc03}}\\
\texttt{imf} & 1 (Chabrier) \\
\texttt{metallicity} [solar] & 0.02 \\
\texttt{seperation\_age} [Gyr] & 0.01 \\
\hline
\multicolumn{2}{c}{\texttt{nebular}}\\
\texttt{logU} & -3.0 \\
\texttt{f\_esc} & 0.0 \\
\texttt{f\_dust} & 0.0 \\
\texttt{lines\_width} [km s$^{-1}$] & 300.0 \\
\hline
\multicolumn{2}{c}{\texttt{dustatt\_modified\_starburst}}\\
\texttt{E\_BV\_nebular} [mag] & [0,1.0], $\delta=0.1$\\
\texttt{E\_BV\_factor} & 0.44 \\
\texttt{uv\_bump\_wavelength} [nm] & 217.5 \\
\texttt{uv\_bump\_width} [nm] & 35.0 \\
\texttt{uv\_bump\_amplitude} & 0.0, 1.5, 3.0 (Milky Way) \\
\texttt{powerlaw\_slope} & [-0.5, 0.0], $\delta=0.1$  \\
\texttt{Ext\_law\_emission\_lines} & 1 (Milky Way) \\
\texttt{Rv} & 3.1 \\
\texttt{filters} & B\_B90, V\_B90, FUV \\
\hline
\multicolumn{2}{c}{\texttt{dl2014}}\\
\texttt{qpah} & 0.47, 2.50, 4.58, 6.63 \\
\texttt{umin} & 0.10, 0.25, 0.50,\\
& 1.0, 2.5, 5.0, 10.0, 25.0 \\
\texttt{alpha} & 2.0 \\
\texttt{gamma} & 0.001, 0.00199526, \\
& 0.00398107, 0.00794328,\\ 
& 0.01584893, 0.03162278,\\
&0.06309573, 0.12589254,\\
&0.25118864, 0.50118723 \\
\hline
\multicolumn{2}{c}{\texttt{restframe\_parameters}}\\
\texttt{beta\_calz94} & False \\
\texttt{D4000} & False \\
\texttt{IRX} & False \\
\texttt{EW\_lines} & 500.7/1.0 \& 656.3/1.0 \\
\texttt{luminosity\_filters} & FUV \& V\_B90 \\
\texttt{colours\_filters} & FUV-NUV \& NUV-r\_prime \\
\hline
\multicolumn{2}{c}{\texttt{redshifting}}\\
\texttt{redshift} & 0 \\
\enddata
\end{deluxetable}

\subsection{[CII]~158~$\mu$m and dust}
Multiple studies have suggested that the [CII]~158~\micron\ emission line strength could be used as a proxy for star formation rates \citep[e.g.][]{Boselli2002, HerreraCamus2015, DeLooze2014}.  This is due to the role [CII] plays as a primary cooling channel for PDRs, where heating is mainly driven by radiation from young stars.  Therefore, as long as the conditions within PDRs are near thermal equilibrium, heating from young stars should be balanced by cooling from far--infrared emission lines, like [CII].  One way to test this proposition is to compare [CII] emission to other known tracers of star formation, like indicators of dust heating.  As dust absorbs UV light emitted by young stars and re--emits infrared light, investigating how changing dust properties influences the strength of the [CII] line helps clarify how [CII] emission is tied to the energy balance in the ISM.  In this section, we will compare the [CII] emission to the amount of attenuated UV light (determined using SED fitting), the far--infrared luminosity, and the strength of the PAH emission.  All of these comparisons help to illuminate how energy is being transformed throughout the ISM.  In addition, [CII]/UV Attenuation, [CII]/FIR, and [CII]/PAH have all been proposed as tracers of photoelectric heating efficiency \citep{Kapala2017, Croxall2012}.  By comparing the local measurements we obtain with FIFI-LS to archival global [CII] measurements, we can also see how the energy balance is a function of local conditions.


\subsubsection{[CII] and UV attenuation}
\begin{figure}[!t]
     \centering
    \includegraphics[width=0.49\textwidth]{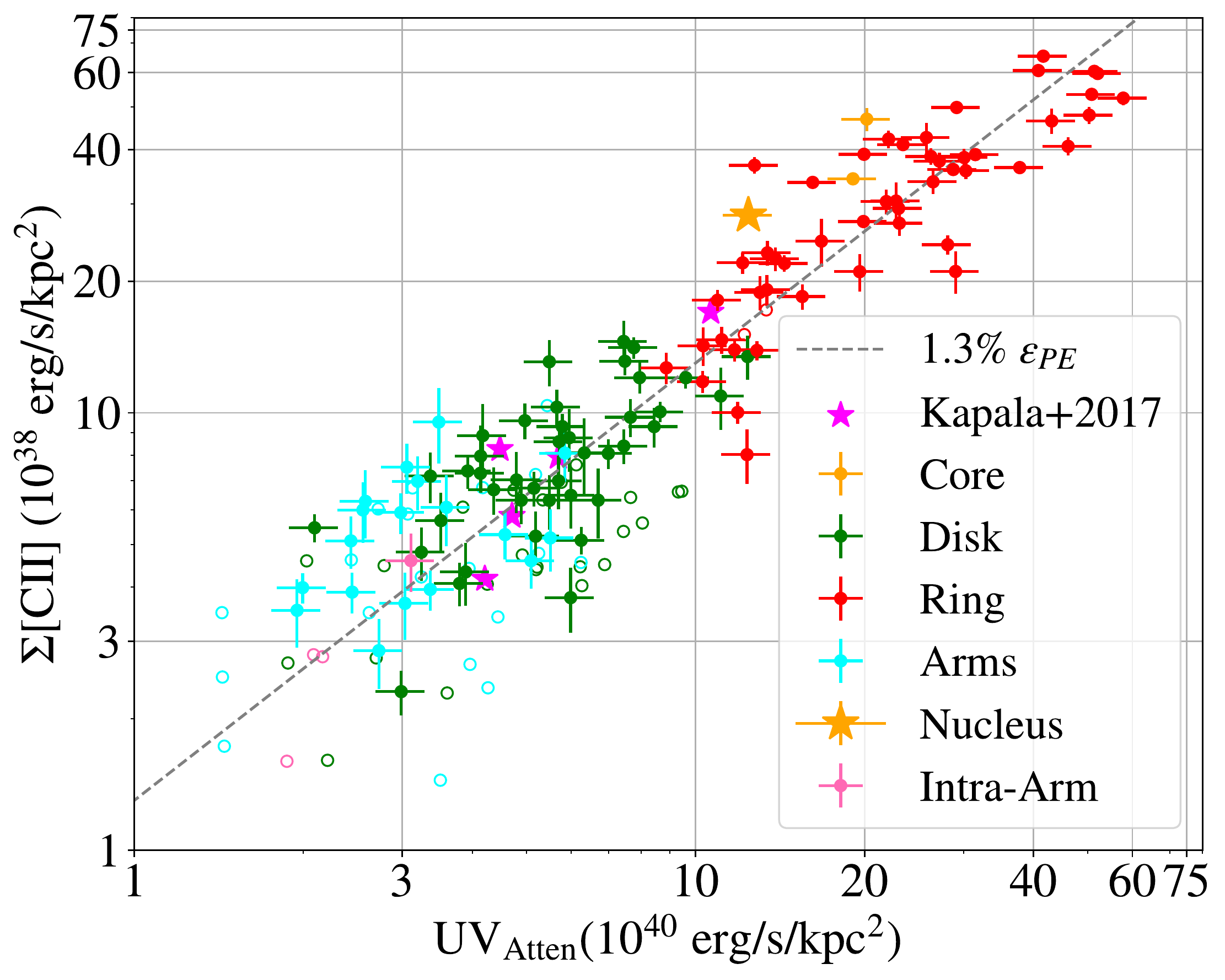}
    \caption{The UV$_{\rm{Atten}}$ values determined with CIGALE plotted against the [CII] surface brightness measurements ($\Sigma$[CII]). Points are color coded according to their environment. The empty circles show points with SNR$<$5. For comparison, the data from the \citet{Kapala2017} study of M31 is plotted as magenta stars. The line shows the predicted $\Sigma$[CII] for an ISM with a photoelectric heating efficiency of 1.3\%.}
    \label{fig:UVAtten}
\end{figure}

The attenuated UV emission (UV$_{\rm{Atten}}$) is estimated based on the results from the CIGALE SED fits (see Section~\ref{sec:SED} for more information on SED fitting).  Using the \citet{BruzualCharlot2003} stellar population models and the CIGALE \verb|dustatt_modified_starburst| module \citep[based on the][attenuation law]{Calzetti2000} the CIGALE program determines the attenuated emission from the two stellar populations used to create the optimal SED model for the inputted fluxes.  UV$_{\rm{Atten}}$ is then computed by integrating the sum of the two stellar attenuation models from $\lambda = 91.2$~\micron\ to  $\lambda = 206.6$~\micron.  Finally, the UV$_{\rm{Atten}}$ values are divided by the deprojected area of the individual regions to obtain the surface brightness of attenuated UV emission. The UV$_{\rm{Atten}}$ values are plotted against the [CII] surface brightness measurements ($\Sigma$[CII]) in Figure~\ref{fig:UVAtten}. 

\begin{figure*}[!t]
     \centering
    \includegraphics[width=0.75\textwidth]{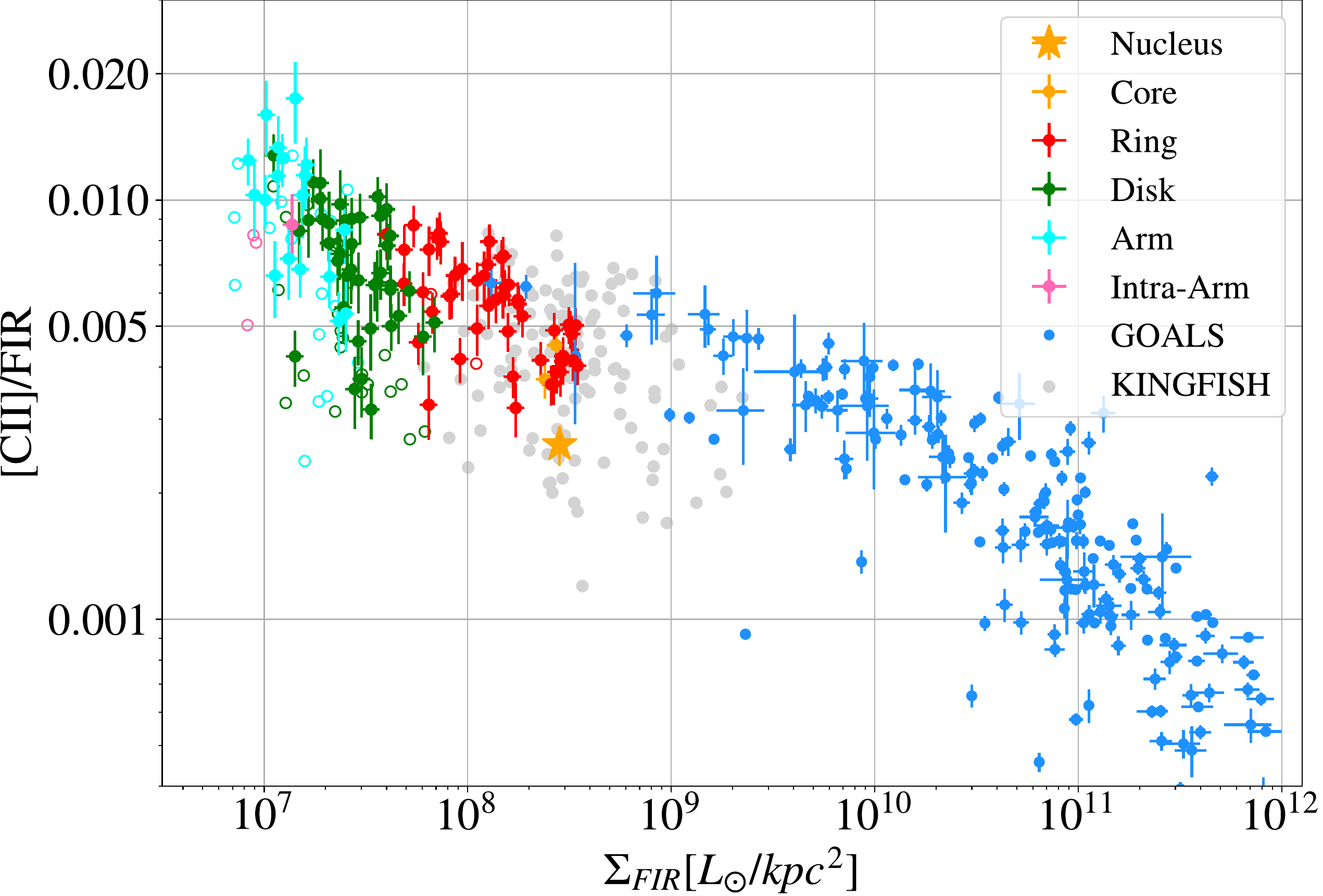}
    \caption{[CII]/FIR values plotted as a function of FIR surface brightness ($\Sigma_{FIR}$ for the regions in NGC~7331 separated by environment). Filled circles have [CII] detection with SNR greater than 5, otherwise they are empty. Additionally, data from the GOALS sample of LIRGS  \citep{DiazSantos2017} and KINGFISH \citep{Sutter2019}  are plotted for comparison.}
    \label{fig:CIIFIR}
\end{figure*}
In addition to the data from this work, Figure~\ref{fig:UVAtten} contains the data from the Survey of Lines in M31 \citep[SLIM, ][]{Kapala2017}.  This survey measured the [CII] emission from five regions along the plane of M31 using the PACS instrument on board \textit{Herschel}.  In order to compare the data from the SLIM survey to NGC~7331, the UV$_{\rm{Atten}}$ values were recomputed using the photometry from \citet{Kapala2017} and the SED modeling method described in this work.  As the SED models used in this work differ from those used in \citet{Kapala2017}, we find systematically higher values of UV$_{\rm{Atten}}$ than those found in \citet{Kapala2017}.  This indicates the model--dependence of UV$_{\rm{Atten}}$ measurements made using SED fitting and therefore the value of the photoelectric heating efficiency derived using this method. 

To determine how galaxy environment effects the relationship between $\Sigma$[CII] and UV$_{\rm{Atten}}$, we divide the data from NGC~7331 into five different environments, based on the locations shown in Figure~\ref{fig:SEDfits}.  Data from regions surrounding the core of the galaxy are colored orange (with the region centered on the nucleus marked with a star), data from within the molecular ring are red, data from within the flocculant disk are green, data from the five spiral arms are cyan, and data from the intra--arm disk are pink.  Although all data points follow the same general trend line with a slope of 0.013 (representing a photoelectric heating efficiency of 1.3\%), there are clear trends between the different environments.  The three data points from the core all lie above the trend line, suggesting either different heating properties or additional [CII] emission from outside of PDRs.  The data from the molecular ring follows the trend line with the least amount of scatter, which implies that the [CII] emission from the molecular ring is well--balanced by heating from attenuated UV emission.  The data from the spiral arms and the inter--arm disk cover similar parameter space, with both lower $\Sigma$[CII] and UV$_{\rm{Atten}}$ than the measurements within the ring and nucleus, and large scatter about the trend line.  This suggests that the [CII] emission from the more diffuse disk is less--closely tied to the UV heating than the [CII] emission from the inner parts of NGC~7331.

\begin{figure*}
     \centering
    \includegraphics[width=0.8\textwidth]{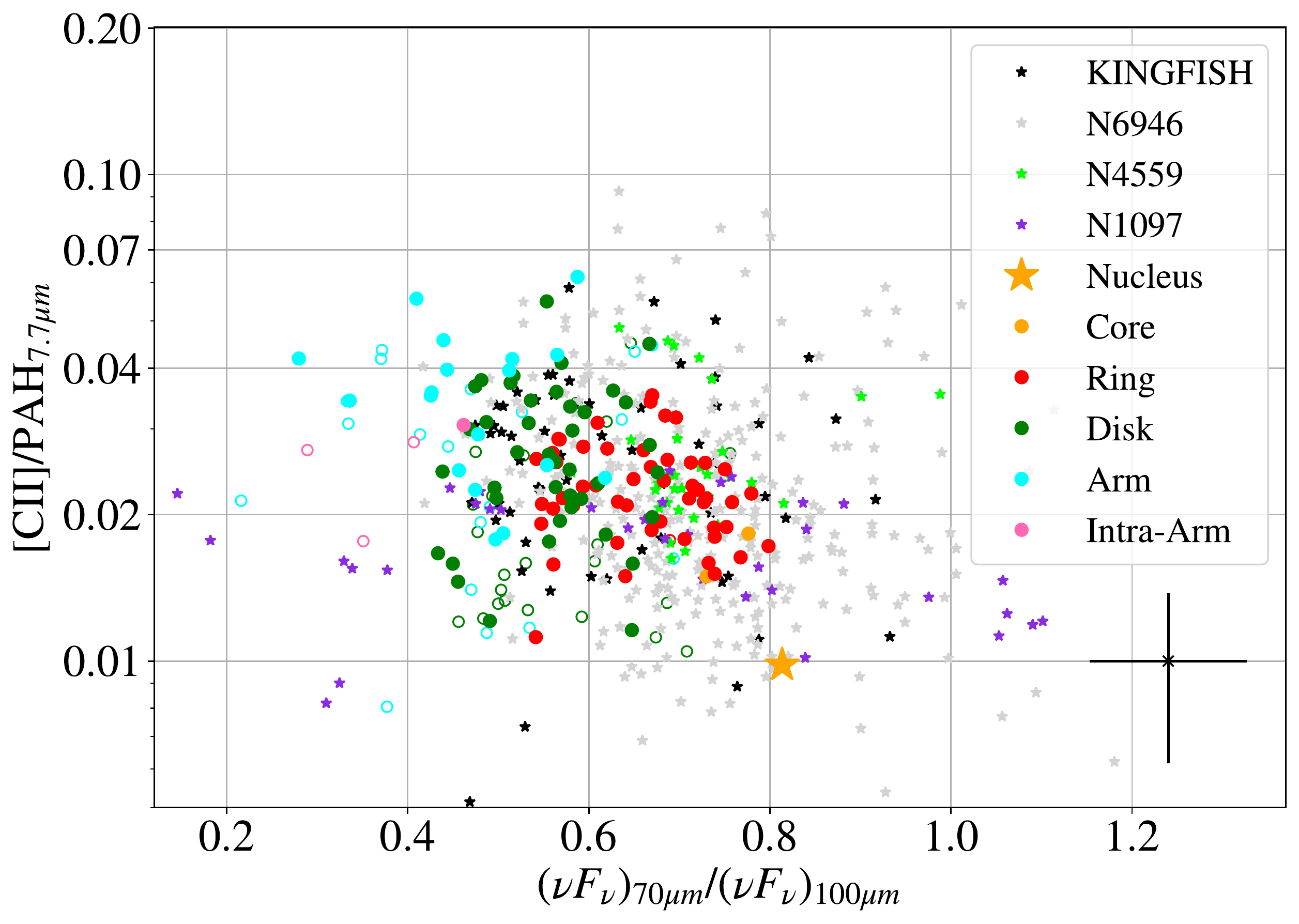}
    \caption{[CII]/PAH values versus the $70\mu m/100\mu m$ color for regions in NGC~7331 are compared to measurements in other galaxies: NGC 6946~\citep{Bigiel2020}, KINGFISH~\citep{Sutter2019}, our re-analysis of NGC~4559 and NGC~1097 published by \citet{Croxall2012}. Filled circles have [CII] detection with SNR greater than 5, otherwise they are empty. The representative error is shown under the legend. }
    \label{fig:CIIPAH}
\end{figure*}

\subsubsection{[CII] and total infrared luminosity}

Another way to measure the photoelectric heating efficiency is by comparing the [CII] emission to the far infrared emission \citep{Malhotra2001}.  As a majority of the FIR emission comes from warm dust, it is a good tracer of heating within the ISM.  Measurements of the ratio of [CII] line emission to far--infrared luminosity have lead to the discovery of the so--called ``[CII] deficit,'' characterized by the decreasing trend in [CII]/FIR as a function of far--infrared color \citep{DiazSantos2017, Croxall2017, Sutter2019}, infrared luminosity \citep{Smith2017, Luhman2003}, and the ratio of infrared luminosity to H$_2$ gas mass \citep[$L_{\rm{IR}}/M_{\rm{H}_2}$,][]{GraciaCarpio2011}.  Understanding the cause of this decrease is of utmost importance for determining the utility of [CII] line emission measurements.  

The [CII]/FIR values for regions in the nucleus, ring, arms, and disk of NGC~7331 are plotted as a function of FIR surface brightness ($\Sigma_{FIR}$) in Figure~\ref{fig:CIIFIR}.  The FIR luminosity was computed using the CIGALE SED fits (described in Section~\ref{sec:SED}) and integrating the modeled luminosity from 8--1000~\micron.  The surface brightness was then computed by dividing this value by the deprojected surface area of each region.  In addition to the data from NGC~7331, the values of [CII]/FIR from the Great Observatories All-sky LIRG Survey \citep[GOALS][]{DiazSantos2017} sample are displayed as blue points and normal, local galaxies from KINGFISH are plotted as grey points \citep{Sutter2019}.  The [CII] emission of the KINGFISH galaxies has been corrected using the same factor found for the ratio between {\sl Herschel} and SOFIA data for the NGC~7331 (see Section~\ref{sec:PACSvsFIFI}) since the observation technique used is similar.  The measurement of the nuclear region of NGC~7331 included in the KINGFISH study falls among our measurements of the core of NGC~7331, likely due to the larger aperture size used in \citet{Sutter2019}.  A more accurate reanalysis of the KINGFISH data will be presented in a future paper.

It is clear that the individual environments within NGC~7331 behave differently.  The outer disk and spiral arms, shown as green, pink, and cyan points, have the lowest $\Sigma_{FIR}$ and the highest [CII]/FIR values, clearly extending the trend observed in the GOALS sample to less-IR luminous regions.  The values from regions within the molecular ring, displayed as red points, form a bridge between the data from the disk of NGC~7331 and LIRGS in the GOALS study, with higher $\Sigma_{FIR}$ than found in the disk and a modest decreasing trend in [CII]/FIR with increasing FIR surface brightness.  The regions surrounding the core of NGC~7331, displayed as orange points, have very low [CII]/FIR values with $\Sigma_{FIR}$ properties similar to those within the ring.  This variation in behavior shows that local observations of the [CII] deficit extend the global trends observed in LIRGs and ULIRGs to a less FIR--luminous regime.

\begin{figure}[!t]
     \centering
    \includegraphics[width=0.45\textwidth]{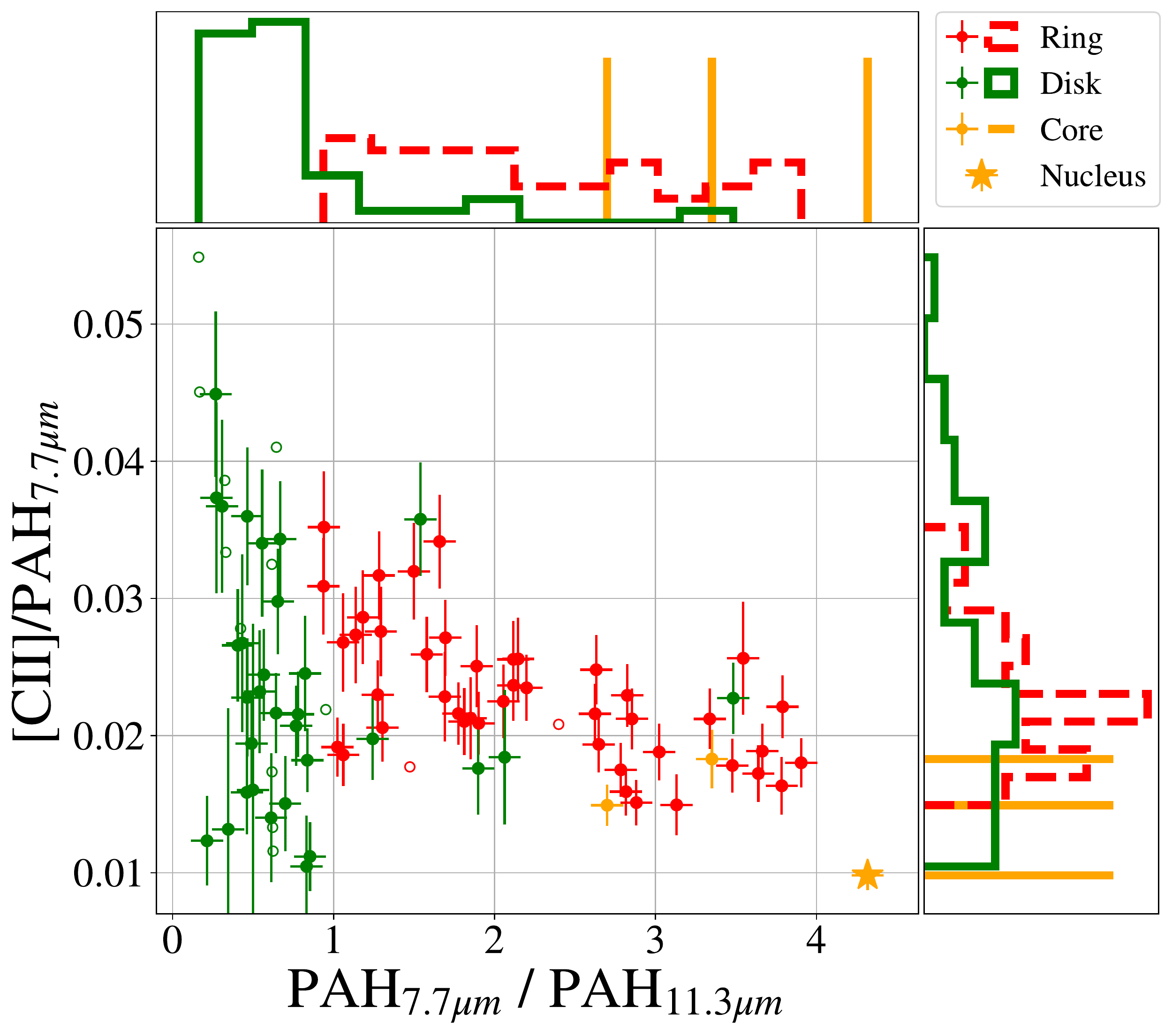}
    \caption{[CII]/PAH values plotted as a function of the ratio of the strength of the PAH features at 7.7 and 11.3~$\mu$m, widely regarded as an indicator of the fraction of charged PAH grains. We find clear differences between PAH$_{7.7\mu m}$ / PAH$_{11.3\mu m}$ values for the disk and ring, as shown by the histogram plotted above the main figure.  In addition, we find a slight decreasing trend in [CII]/PAH as a function of PAH$_{7.7\mu m}$ / PAH$_{11.3\mu m}$, indicating that increasing PAH charge could be effecting the photoelectric heating efficiency, and therefore lowering the [CII]/PAH values in hotter ISM conditions.}
    \label{fig:PAHCharge}
\end{figure}

\subsubsection{[CII] and PAH emission}

The final method we use to track the photoelectric heating efficiency across NGC~7331 is to compare the [CII] emission to the emission from polycyclic aromatic hydrocarbons (PAHs).  PAHs are considered to be a significant source of free electrons in the ISM due to the low energy needed to photoeject electrons from these molecules \citep{Tielens2008}.  This leads to PAH emission being closely--tied to the heating in the ISM.  For this analysis, PAH emission is inferred from IRAC channel 4 data corrected for stellar emission by finding the difference between the flux seen by IRAC4 and the unattenuated stellar flux in this band determined using the CIGALE SED fits.  To find the modeled stellar flux within the IRAC4 band, the summation of the luminosity of the young and old stellar populations in the CIGALE models was multiplied by the transmission curve for the IRAC band 4 filter.  The stellar flux is then subtracted from the observed IRAC band 4 flux to determine the strength of the 7.7~\micron\ PAH feature.  For comparison, the data from NGC~1097 (barred spiral galaxy) and NGC~4559 (intermediate spiral galaxy) are plotted as purple and green points, respectively. These two galaxies were used by \citet{Croxall2012} to study the relationship between [CII] and PAH emission. However, the values presented in this study are affected by inaccurate calibration of the [CII] emission which is evident when comparing the SOFIA based results from \citet{Bigiel2020} on NGC~6946 to \citet{Croxall2012} values as done in \citet{Fadda2021} (Fig. 7). 
We recomputed the [CII] values of NGC~1097 and NGC~4559 using the latest PACS calibration and limiting our study to the data taken in standard chop-nod mode. The rest of the observations were done in wavelength switching mode, a mode later abandoned because it is difficult to calibrate. As shown in Figure~\ref{fig:CIIPAH}, these two galaxies occupy the same locus as NGC~6946 \citep{Bigiel2020} and our galaxy, as well a compilation of measurements from KINGFISH galaxies \citep{Sutter2019}.  The measurement of the nuclear region from \citet{Sutter2019} falls among our measurements of the core of NGC~7331.  Further information about the necessary corrections to the KINGFISH data will be presented in a future paper.


\begin{figure*}[!t]
\begin{center}
\includegraphics[width=0.95\textwidth]{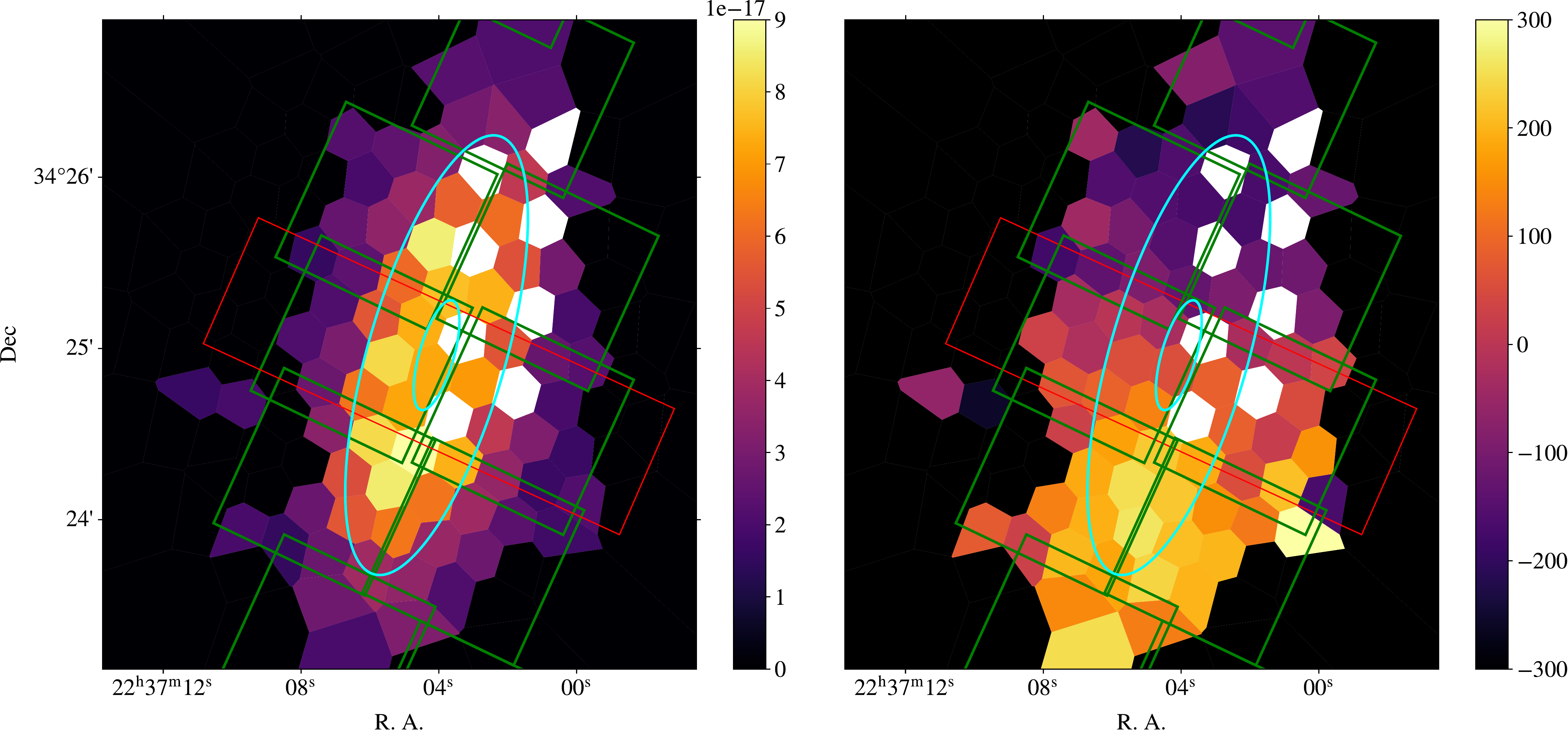}
\end{center}
\caption{
SPIRE intensity (left) and velocity (right) maps for the [NII]~205~$\mu$m line. The intensity is in units of 10$^{-17}$~W/m$^2$ and the velocity is in km/s with respect to the system velocity of 800~km/s. The maps are shown as Voronoi tessellations, with tiles centered on the positions of the SPIRE detectors. The white tiles correspond to the two dead detectors of SPIRE in the four jiggle positions. For each detector, the unapodized spectrum has been fitted with the convolution of a Gaussian and a sinc function. Contours have the same meaning as in Figure~\ref{fig:coverage}.
}
\label{fig:SPIRE}
\end{figure*}


\begin{figure*}[!t]%
    \centering
    \includegraphics[width=7cm]{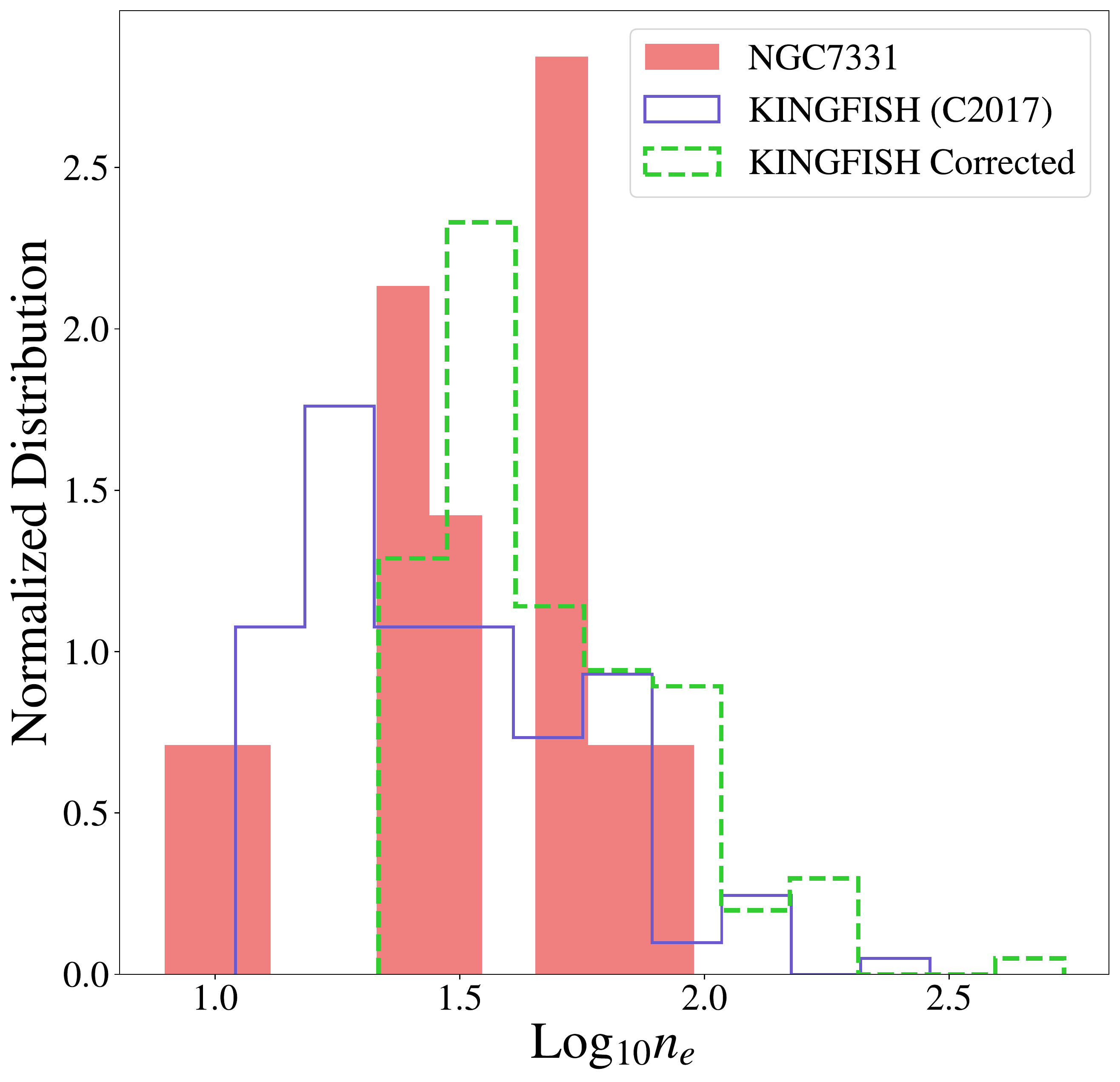}%
    \qquad
    \includegraphics[width=7cm]{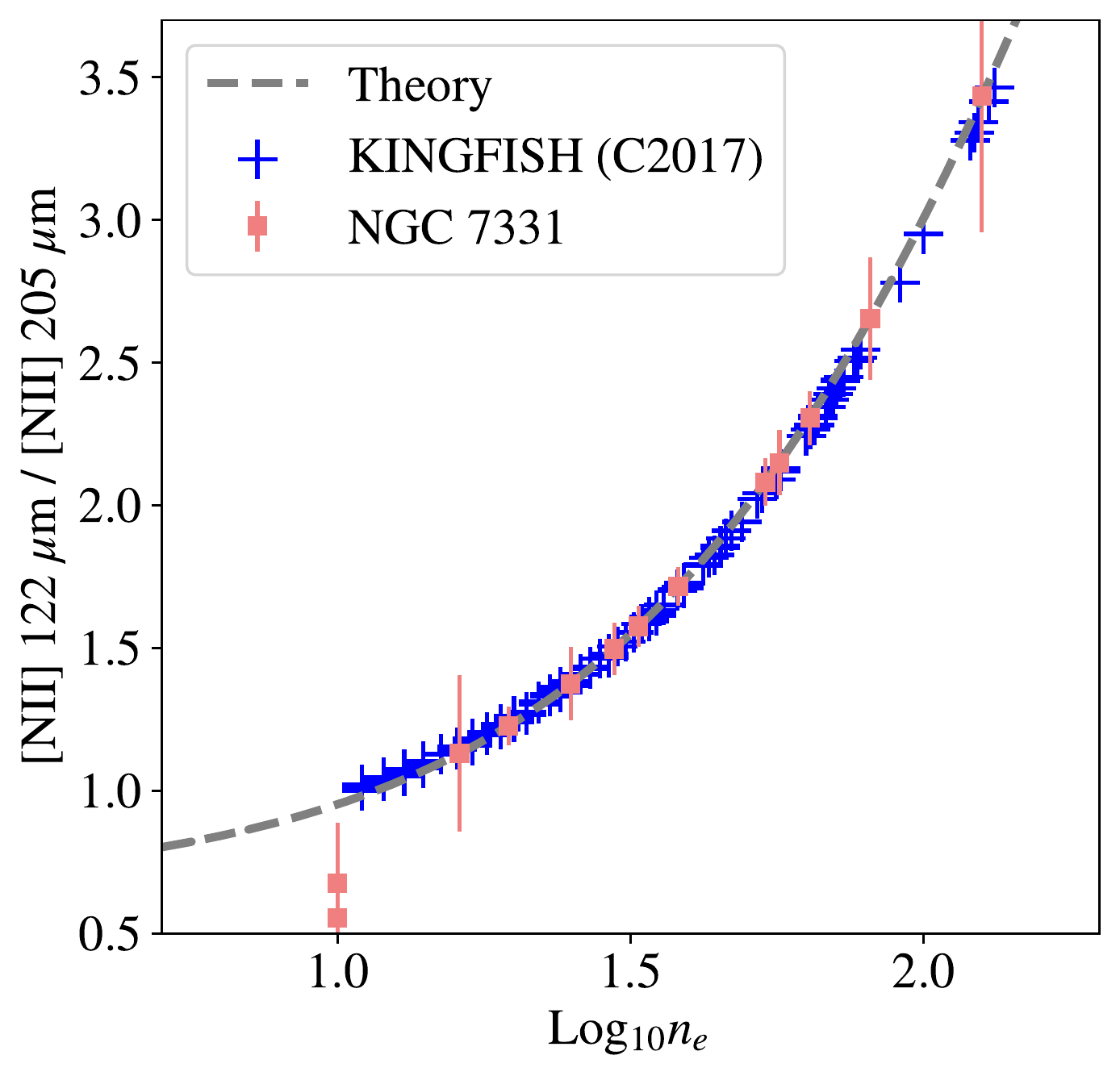}%
\caption{\textit{Left:} The electron density values in the SPIRE detectors with corresponding PACS detections computed using [NII]122/205 is shown as a red--bar histogram.  For comparison, electron densities computed in \citet{Croxall2017} for regions in the local galaxies in the KINGFISH survey are shown as a solid blue line. The offset between the values found in \citet{Croxall2017} and NGC~7331 is due to differences in the calibration of the PACS [NII]~122~\micron\ line (see Section~\ref{sec:PACS} for more information). The dashed--green line  shows the distribution of $n_e$ values obtained by first scaling the [NII]~122~\micron\ fluxes from \citet{Croxall2017} by 20\%, the ratio between the fluxes determined with the two different reductions, and then re--computing $n_e$.
\textit{Right:} The measurements of [NII]~122/205 plotted against the derived $n_e$ values for NGC~7331 (red squares) and the KINGFISh data (blue crosses).  The theoretical relationship between [NII]~122/205 and $n_e$ used is shown as a gray dashed line.} 
\label{fig:n_e}
\end{figure*}

One possible explanation for the slight decreasing trend we observe between the [CII]/PAH and increased $(\nu F_{\nu})_{70} / (\nu F_{\nu})_{100}$ seen in Figure~\ref{fig:CIIPAH} is that increasing fraction of charged PAH grains could decrease the photoelectric efficiency in hotter PDRs, and therefore lower the [CII] emission strength \citep{Malhotra2001, Croxall2012}.  The effects of grain charging can be tested by examining the ratio of the strengths of the 7.7~\micron\ 11.3~\micron\ PAH features (PAH~7.7/11.3), as increased PAH~7.7/11.3 indicates a higher fraction of charged PAH grains \citep{Draine2021}. We determine the strength of the 11.3~\micron\ feature using the ISOCAM~LW8 band fluxes and the results of the CIGALE SED fits.  Following a similar method as was used to determine the 7.7~\micron\ PAH feature strength, the unattenuated stellar fluxes within the ISOCAM~LW8 band determined from the SED fits were subtracted from the observed ISOCAM data.  In addition, due to the longer wavelength of the 11.3~\micron\ feature, we also subtract the modeled thermal dust emission from the ISOCAM fluxes, to leave only the emission from the 11.3~\micron\ PAH feature.  The \CII/PAH values are plotted against the PAH~(7.7/11.3) values in Figure~\ref{fig:PAHCharge}.  As the ISOCAM data covers a smaller area than the FIFI-LS data, only the regions with full ISOCAM coverage are included.  For this reason, none of the regions from within the spiral arms are plotted in this figure.

As shown in Figure~\ref{fig:PAHCharge}, there is a slight decrease in the [CII]/PAH ratio with increasing PAH~7.7/11.3.  This is similar to observations from \citet{Croxall2012}, and suggests that increasing PAH charge could be playing a role in the decreasing [CII]/PAH values for regions with higher infrared colors.  We note that the region centered on the nucleus is likely effected by the presence of a dust--enshrouded AGN, and therefore the anomalously low [CII]/PAH and high PAH~7.7/11.3 values measured for this region are likely due to the extreme conditions of the ISM surrounding the buried AGN.

\subsection{[CII] from the Ionized ISM}
One of the challenges of using the [CII] line to measure properties of the ISM is the ubiquity of singly--ionized carbon.  As carbon has a relatively low ionization potential of 11.3~eV, singly--ionized carbon is the predominant phase of carbon in both the ionized and neutral ISM \citep{Bennet1994, Pineda2013, HerreraCamus2015}.  In order to further understand the origins of the [CII] emission in NGC~7331, the ratio of the [CII]~158~\micron\ line and the [NII]~205~\micron\ line can be used to estimate the fraction of the [CII] emission originating in ionized phases of the ISM \citep{Parkin2013, Croxall2017, DiazSantos2017, Sutter2019}.  The unique origin of the [NII] lines allows for models of [CII]~158~\micron/[NII]~205~\micron\ to be used to estimate the fraction of the [CII] emission that originates co--spatially with the [NII] emission in ionized phases of the ISM and the fraction that originates in the neutral phases where little to no ionized nitrogen is present.  The 205~\micron\ line of singly--ionized nitrogen is the preferred line for making these measurements due to the similarity between the critical densities of the [NII]~205~\micron\ line and the [CII]~158~\micron\ line in collisions with electrons \citep[$\sim 44$~cm$^{-3}$ and $\sim 45$~cm$^{-3}$ for temperatures of the warm ionized medium of T$\sim$8000~K, respectively,][]{Oberst2006}, making the theoretical ratio only weakly dependant on density conditions.  To determine this fraction, we first reduce the available SPIRE 205~\micron\ data for a direct comparison to the \CII.  In addition, we determine the electron number density in the ionized ISM using the ratio of the 122~\micron\ and 205~\micron\ [NII] lines where data is available to more accurately determine the theoretical \CII/[NII]~205~\micron\ ratio used to predict the fraction of \CII\ from the ionized ISM.

\begin{figure*}[!t]
     \centering
    \includegraphics[width=0.7\textwidth]{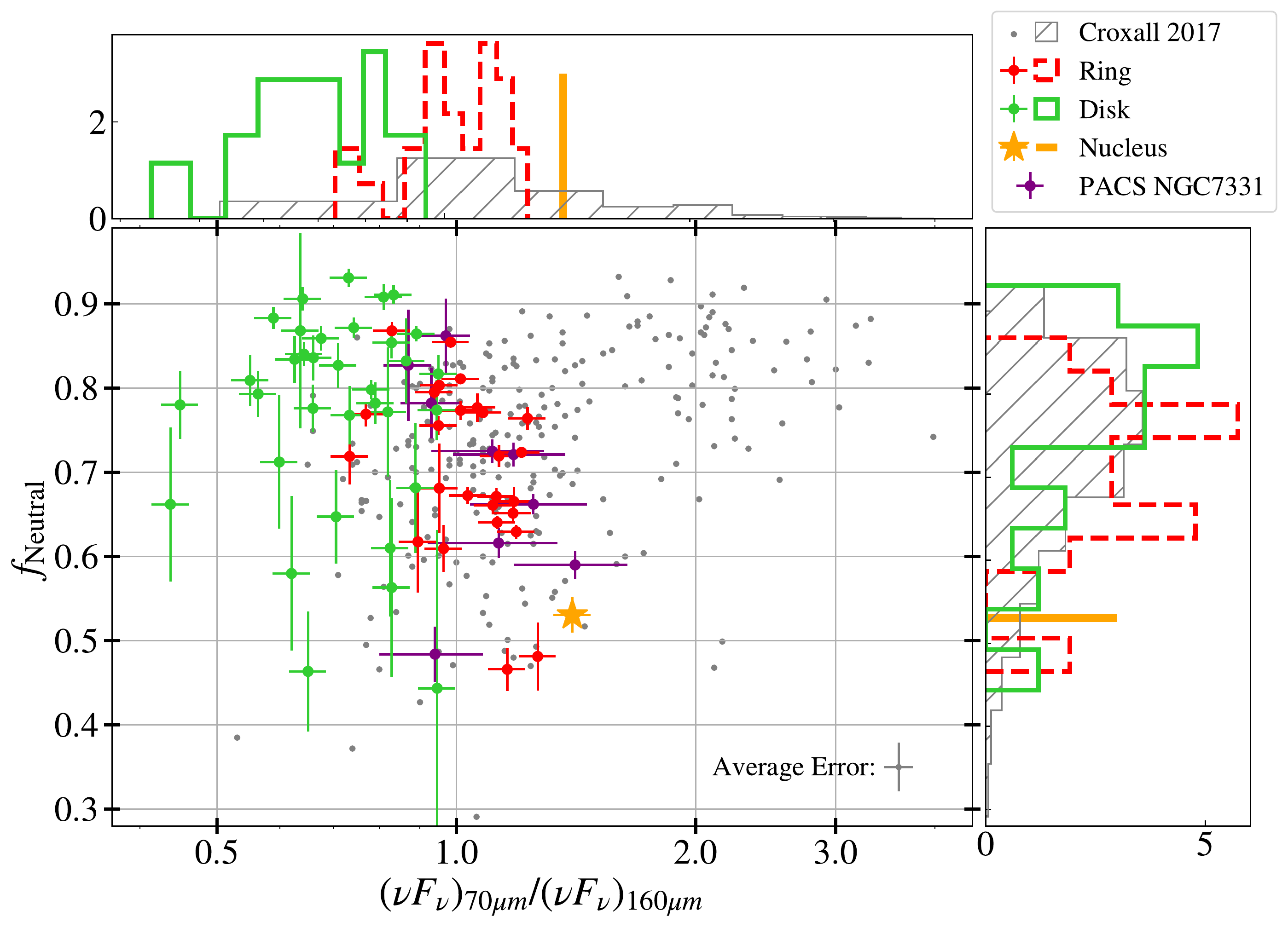}
     \caption{Fraction of [CII] emission from the neutral phases of the ISM plotted as a function of  $(\nu F_{\nu})_{70 \mu m} / (\nu F_{\nu})_{160\mu m}$.  NGC~7331 data from this work (color-coded according to the environment) are compared to data from \citet{Croxall2017} (gray points). NGC~7331 data points from the \citet{Croxall2017} paper are highlighted in purple to emphasize the good agreement between the PACS data used in \cite{Croxall2017} and the SOFIA data used in this work.  In addition, the histogram on the top of the plot shows the differences in FIR color distributions between disk, ring, and nucleus. The histogram on the right shows the good agreement with the \citet{Croxall2017} distribution of $f_{\rm{Neutral}}$ values.
    }
    \label{fig:fneut_comps}
\end{figure*}

\subsubsection{[NII]~205~$\mu$m SPIRE map}
In order to determine the fraction of the \CII\ emission originating in different phases of the ISM, we use the [NII]~205~\micron\ SPIRE map.  Since SPIRE uses Fourier transform spectroscopy, the natural results of the detectors are interferograms. The Herschel pipeline also produces spectra which are the result of the convolution of the original spectra with an apodizing function. Such a function reduces the size of the sidelobes and transform the profile of the lines into Gaussians. Unfortunately, this process broadens the line profile resulting in a loss of spectral resolution. Moreover, since the shape of the line is not exactly Gaussian, the convolution also introduces a bias in the total flux which, in the case of SPIRE, amounts to approximately 5\% according to the Herschel manual. 
For these reasons, we measured the [NII]~205~\micron\ emission lines for each single detector
by fitting the line in the unapodized spectra. Unresolved lines can be fitted with a {\sl sinc} function. However, at 205~\micron\ the spectral resolution is around 1020 and typically galaxy lines are broad. So, rather than using a simple {\sl sinc} function, we adopted a convolution of a Gaussian and a {\sl sinc} function which describes the line profile in a more general way. For the fit, we adopted the robust analytical formulation of the convolution proposed by \citet{Martin2016} which makes use of the Dawson integral instead of the error function ({\sl erf}), the standard solution adopted in the Herschel software. As shown by \citet{Martin2016}, the formulation used in the Herschel software cannot be computed too far from the central lobe because of overflow errors, while the formula with the Dawson integral, which is mathematically equivalent to the formulation with the {\sl erf} function, can easily fit the function far away from the central lobe.  By comparing the fluxes of the 205~\micron\ line estimated with this function to those estimated using only the {\sl sinc} function, we see that simple fits with {\sl sinc} functions miss approximately 20\% of the flux. So, considering the convolution of the {\sl sinc} function with the Gaussian is, in this case, particularly important. For the fit, we also adopted the maximum optical distance path of 12.56~cm as reported by \citet{Fulton2016} in the case of high resolution spectra (see Table~1 of their paper).
Figure~\ref{fig:SPIRE} shows the intensity and velocity maps of the 205~\micron\ line. We only considered lines detected with signal-to-noise ratios greater than 5. Since the SPIRE detectors uniformly cover the region observed with circles which are slightly overlapping, we decided to use a Voronoi tessellation to display the flux by maintaining the center of the detectors as tile centers. The regions covered by the dead detectors are colored in white in the figure.
The intensity map clearly show the emission from the molecular ring. Unfortunately the four-position dithering of the observation is not sufficient to recover the intrinsic spatial resolution of SPIRE at this wavelength. Only with full spatial sampling (16 dithering positions) would it have been possible to see the ring of dust and molecular gas. The velocity map shown in the right panel of Figure~\ref{fig:SPIRE} clearly shows the rotation of the galaxy.

\subsubsection{Electronic density from [NII] line ratio}
\label{sec:ne}
The electron number density can be determined by comparing the ratio of the [NII]~122~\micron\ and [NII]~205~\micron\ emission lines (hereafter [NII]~122/205).  Derivation of the relationship between [NII]~122/205 and $n_e$ can be found in \citet{HerreraCamus2016}.  This model was used to determine the closest $n_e$ value for the [NII] 122/205 values for the SPIRE detectors with full PACS coverage.  The results of this modelling can be found in the right--hand panel of Figure~\ref{fig:n_e}, where the theoretical curve from \citet{HerreraCamus2016} is shown as a black—dashed line and the values of [NII]122/205 from our sample are shown as red squares.  The data from \citet{Croxall2017} is also shown as blue crosses.  This relationship is ideal for determining the density of the ionized gas as nitrogen has an ionization potential of 14.5~eV, above the 13.6~eV required to ionize hydrogen, meaning singly--ionized nitrogen is primarily found in the ionized ISM.  The electron densities estimated using this method can therefore be used to predict the [CII]/[NII]~205~\micron\ emission ratio.  As this ratio probes the ionized phases of the ISM, a temperature of 8000~K is used for the theoretical estimation of $n_e$.  This ratio is sensitive to electron densities between 10--1000~cm$^{-3}$. For the areas covered by both the SPIRE~205~\micron\ maps and the PACS~122~\micron\ maps, only one has a [NII]~122/205 ratio indicating an electron density outside of this range.  A histogram showing the electron density measurements for the SPIRE detectors within the PACS observations is shown in Figure~\ref{fig:n_e} as red bars.  The relatively sparse PACS coverage resulted in only 12 SPIRE detectors for which $n_e$ could be determined.  This small number of regions, including some with low SNR, is likely the cause of the non--continuous distribution of $n_e$ values in this histogram.  In addition to our $n_e$ measurements, we display the data from \citet{Croxall2017} which includes all the KINGFISH star--forming regions with both [NII] 122 and 205~\micron\ detections, as a solid--blue line histogram. The direct comparison of  of our [NII]~122~$\mu$m measurements with those of \citet{Croxall2017} for NGC~7331 provided evidence a difference in flux of 20\%, as mentioned in Section~\ref{sec:PACS} and further discussed in the appendix Section~\ref{sec:NIIdata}. We investigated the reason of this difference and found it was due to the effect of transients in the PACS data. An in--depth discussion of the findings can be found in the appendix (section~\ref{sec:NIIdata}).  After correcting [NII]~122~\micron\ data from \citet{Croxall2017} for this 20\% scaling factor, the distribution of values from \citet{Croxall2017} compares well with the distribution of $n_e$ values found in NGC~7331, as shown by the green--dashed line histogram in Figure~\ref{fig:n_e}.

\begin{figure}[!t]
     \centering
    \includegraphics[width=0.5\textwidth]{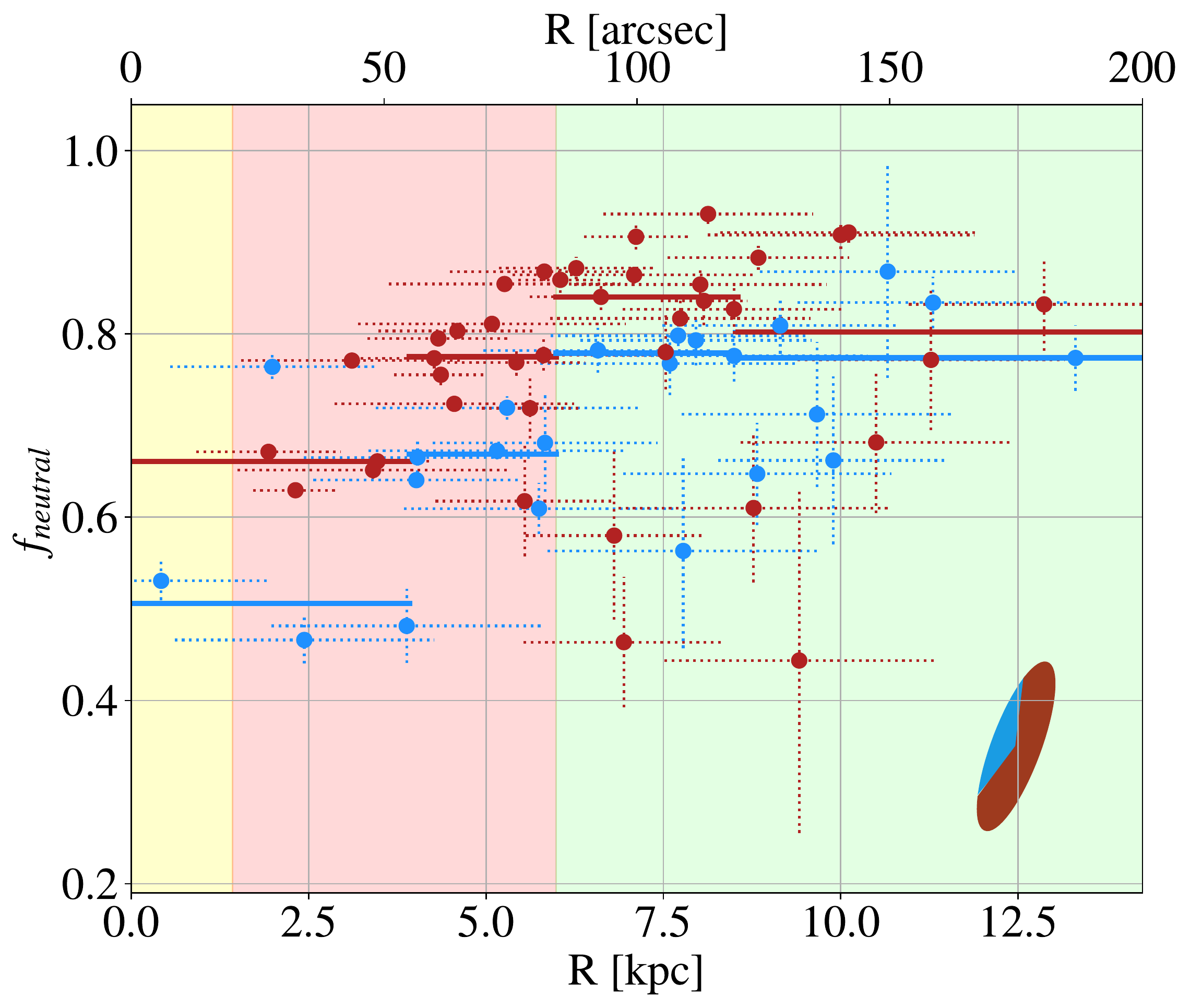}
    \caption{
    Neutral fraction versus deprojected distance from the galaxy center. The color of the points depends on what side they are with respect to the major axis of the ellipse (blue points from the far side, red points from the near side, see shaded ellipse in the right bottom corner for a visualization). The horizontal bar corresponds to the size of the SPIRE beam along the deprojected radius. The solid lines are the median of the values in 4 different intervals. The nuclear, ring, and outer disk regions are shaded in light yellow, red, and green, respectively.
    }
    \label{fig:fneut_radius}
\end{figure}

\begin{figure*}[!t]
\begin{center}
\includegraphics[width=0.68\textwidth]{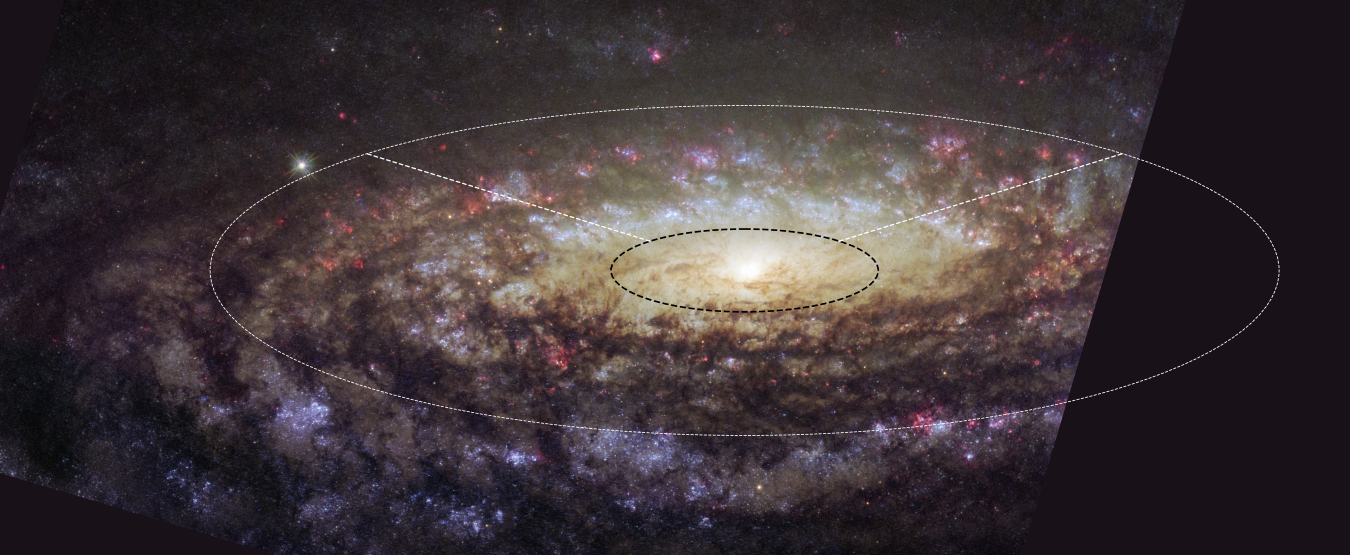}
\includegraphics[width=0.68\textwidth]{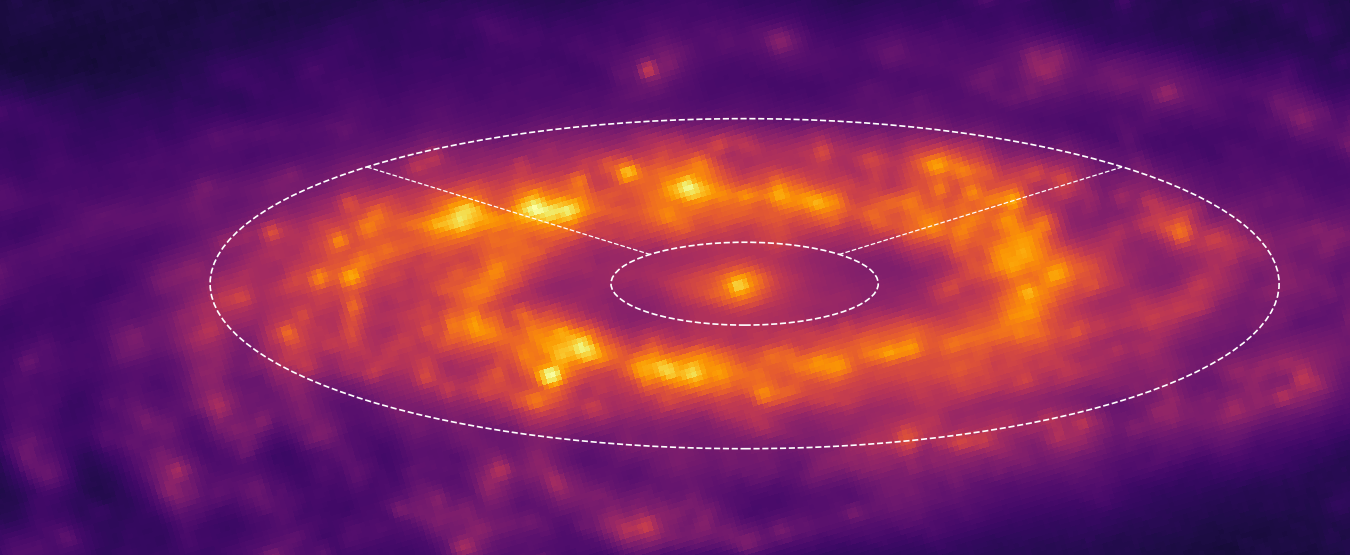}
\includegraphics[width=0.68\textwidth]{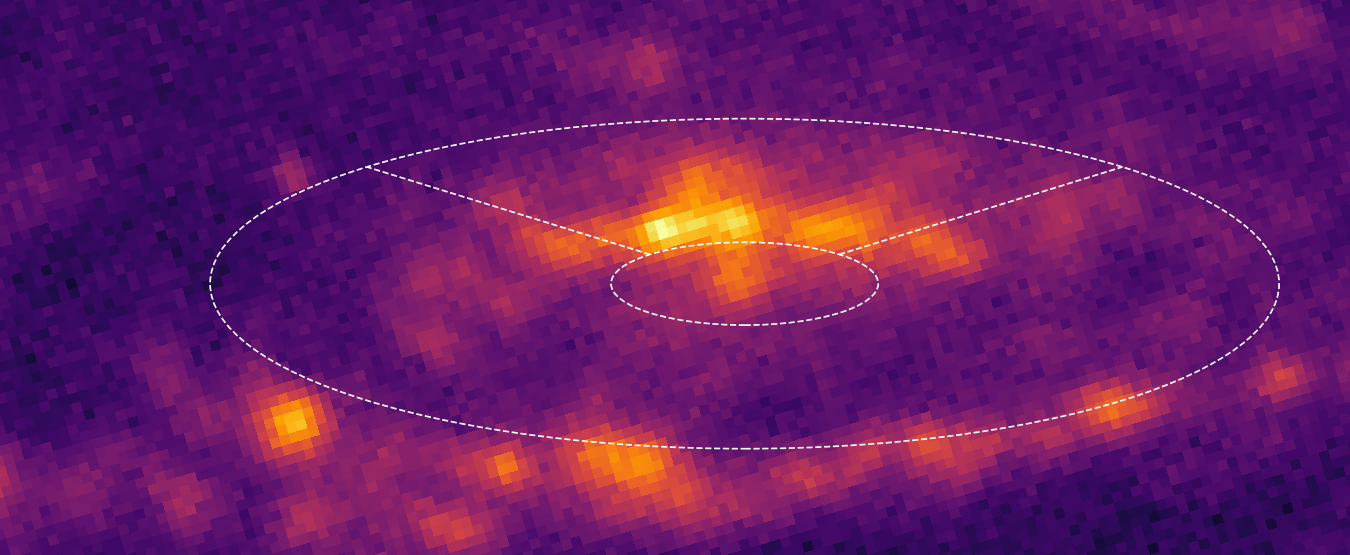}
\end{center}
\caption{
Combined HST images~\textsuperscript{\dag}
(top), IRAC 8~$\mu$m (middle) and GALEX NUV (bottom) images rotated
to align horizontally the major axis of the galaxy. The ring, delimited by two concentrical ellipses, is equally bright at any angle in infrared images. In optical and UV images it appears heavily obscured in the part closer to us (lower sector) and populated by young stars and HII regions on the far side (upper sector).
}
\label{fig:HSTring}
{\footnotesize\textsuperscript{\dag} Credits: ESA/Hubble \& NASA/D. Milisavljevic (Purdue University)}
\end{figure*}

\subsubsection{Origin of the [CII] emission}
\label{sec:fneutral}

We compute the fraction of the \CII\ emission that originates in the neutral ISM ($f_{\rm{Neutral}}$) using the formula:
\begin{equation}
    f_{\rm{Neutral}} = \frac{\rm{[CII]} - R_{\rm{Ionized}}\times\rm{[NII]}205\mu\rm{m}}{\rm{[CII]}}
\end{equation}
where $R_{\rm{Ionized}}$ is the theoretical ratio of \CII/[NII]~205~\micron\ emission, and [CII] and [NII]~205~\micron\ are the fluxes of the two lines.  For the SPIRE detectors that overlapped with the PACS 122~\micron\ detections, $R_{\rm{Ionized}}$ is determined using the electron number density derived using the method described in Section~\ref{sec:ne}.  For the detectors without [NII]~122~\micron\ measurements, an average value of $n_e$ was used.  As the \CII/[NII]~205~\micron\ ratio is only weakly dependant on density, this will not have a significant effect on the accuracy of our derivation of $f_{\rm{Neutral}}$.  The average value of $f_{\rm{Neutral}}$ determined for the regions in NGC~7331 was 75\%, with a range of 42--95\%.  This is similar to what has been found for the star--forming galaxies of the KINGFISH survey \citep[$74\% \pm 8$, ][]{Croxall2017} and the U/LIRGS studied in GOALS \citep[60--95\%][]{DiazSantos2017}. 

To further compare the KINGFISH sample with our NGC~7331 data, Figure~\ref{fig:fneut_comps} shows the $f_{\rm{Neutral}}$ values plotted against the $(\nu F_{\nu})_{70 \mu m} / (\nu F_{\nu})_{160\mu m}$ for both this work (green, red, and yellow circles) and the \citet{Croxall2017} study (gray points).  As the \citet{Croxall2017} study included a PACS [CII] observations of NGC~7331, these data have been highlighted in purple to show the consistency between the analysis presented in our work and the work of \citet{Croxall2017} while also highlighting the expanded parameter space included in the new \CII\ map.  Since the KINGFISH survey focused on the nuclei of galaxies and extranuclear star--forming regions, the points from the KINGFISH data cover a higher range of $(\nu F_{\nu})_{70 \mu m} / (\nu F_{\nu})_{160\mu m}$ values than our NGC~7331 data.  This is expected as our coverage of NGC~7331 is more extended, covering the cooler inter--arm regions not targeted in the KINGFISH survey.  Irregardless of this difference, there is fairly good agreement between the subset of PACS detections within NGC~7331 and the SOFIA detections included in this work, showing that our methods produce similar results to the \citet{Croxall2017} study.   

\begin{figure*}[!t]
    \centering
    \includegraphics[width=0.9\textwidth]{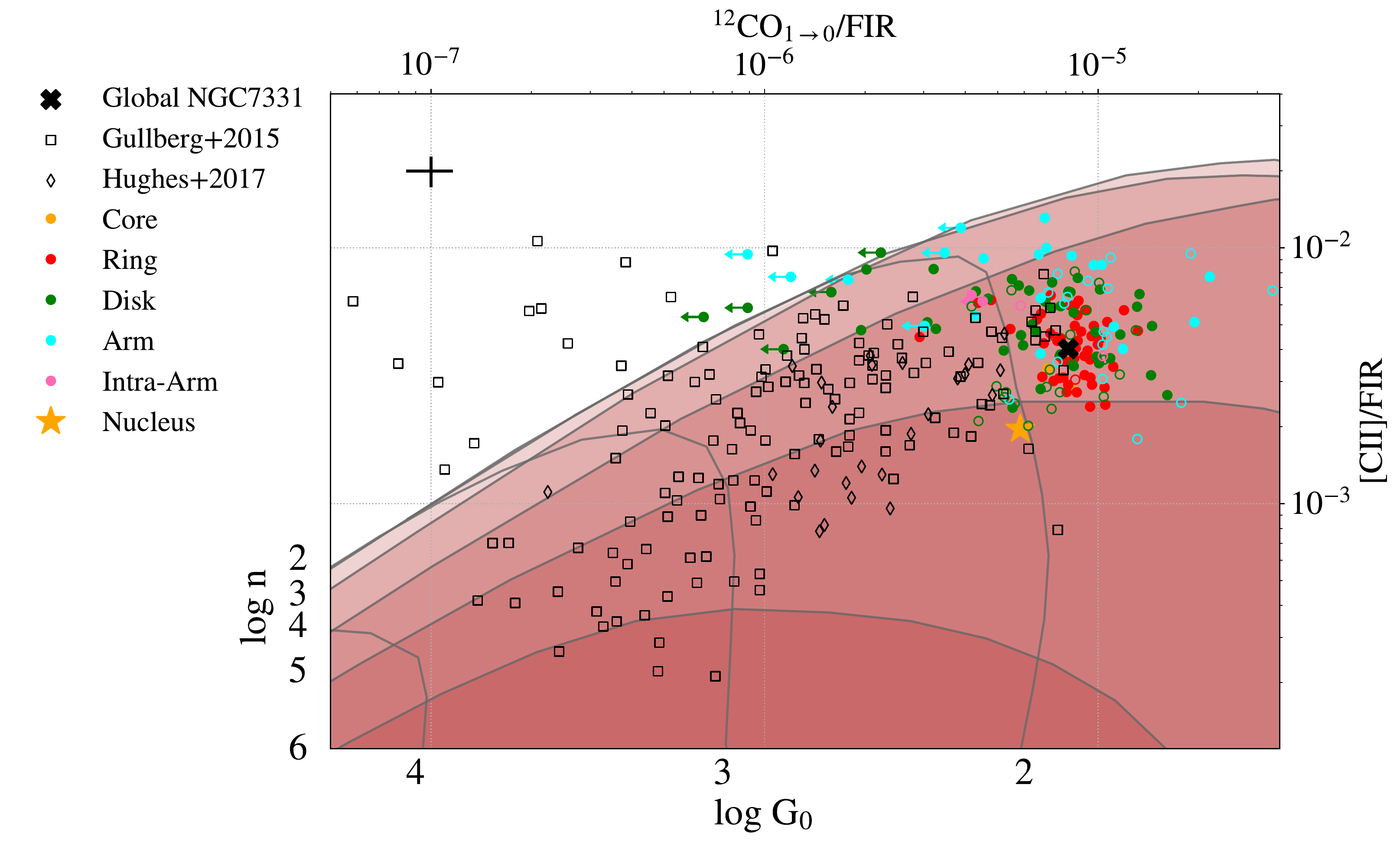}
    \caption{The $^{12}$CO$_{1\rightarrow 0}$/FIR values plotted against the [CII]/FIR values with the PDR models of \citet{Kaufman2006, Pound2008} shown as red shaded regions.  In addition to the data from NGC~7331, archival observations of normal star--forming galaxies from \citet{Hughes2017} and \citet{Gullberg2015} are shown as black--outlined diamonds and squares, respectively.  For comparison, the global measurement of NGC~7331 is displayed as a large black x. Empty circles are apertures with SNR$<$5. Regions with no detectable CO emission are marked as upper limits.  The representative error bar for our data is shown in the upper left corner, scaled to be the same linear size as we would expect at the locus of the points from NGC~7331 for clarity.}
    \label{fig:CIICO}
\end{figure*}

In Figure~\ref{fig:fneut_radius}, $f_{\rm{Neutral}}$ is plotted as a function of galaxy radius.  The radius for each SPIRE detector was determined by deprojecting its distance from the center of NGC~7331, as described in Section~\ref{sec:lumprofile}. The radius is divided into three regions: nucleus, ring, and outer disk shaded in light yellow, red, and green, respectively.  
Using this method, we observe a modest correlation between the fraction of [CII] emission originating in the neutral phases of the ISM and radius, which has a Spearman rank coefficient of 0.38 and a p-value of 0.002. In addition to the radial variation, we color code the data points to show which side of the galaxy they are located. Those on the side closer to the observer are colored brown, while those from the distant quarter are colored blue (see shaded elliptical insert in figure for a visual representation, as well as the dashed outline in Figure~\ref{fig:HSTring}).
There is a marked difference between the distribution of the two populations. Points from the side closer to us have higher neutral fractions than those from the far side of the galaxy. 
If we look at Figure~\ref{fig:HSTring}, we can see that the closest side of the galaxy is obscured in optical wavelengths, while the far side is shining with young white stars and HII regions.  This suggests that viewing perspective can effect what fraction of [CII] emission we observe originates in the neutral phases of the ISM. On the closest side we see the dust--obscured side of the molecular ring and [CII] mainly emitted by neutral gas.  On the other hand, in the far side we see the part of the ring which is illuminated by bright stars and the fraction of [CII] emitted in ionized gas is more important.  Typically, studies of [CII] have targeted galaxies with small inclination angles, especially those viewed face-on.  For example, within the KINGFISH sample, 94\% of the galaxies with some [NII] coverage have inclination angles smaller than that of NGC~7331, and very few edge on galaxies have been mapped at the [NII]~205~\micron\ line \citep[see][for an example]{Hughes2014}.  This analysis of NGC~7331 shows how inclination can alter our view of the [CII] emission, and should be considered when studying high-redshift galaxies for which the inclination angle is unknown.

\subsection{[CII]~158~$\mu$m vs CO}

In addition to using [CII]~158~\micron\ emission as an indicator of ISM heating and cooling properties, it has been suggested that [CII] emission could be used to trace the CO--dark molecular gas, especially in low--metallicity galaxies \citep[e.g.][]{Madden2020}.  To explore this possibility, we plot [CII]/FIR against $^{12}$CO$_{1\rightarrow 0}$/FIR in Figure~\ref{fig:CIICO}.  The $^{12}$CO$_{1\rightarrow 0}$ fluxes were determined by dividing the $^{12}$CO$_{2\rightarrow 1}$ fluxes measured using the HERACLES maps by a factor of 1.2 (see Section~\ref{sec:COdata}).

In order to compare our measurements to PDR models, we multiply the [CII] fluxes by a factor of 0.75, the average value of $f_{\rm{Neutral}}$ found in Section~\ref{sec:fneutral}.  This allows us to estimate the fraction of the [CII] emission that originates in PDRs.  In addition, we increase the $^{12}$CO$_{1\rightarrow 0}$ fluxes by a factor 2 to account for the likelihood that the $^{12}$CO$_{1\rightarrow 0}$ line will become optically thick in dense star--forming regions \citep{Hughes2017}.  

The \CII/FIR are plotted against the $^{12}$CO$_{1\rightarrow 0}$/FIR in Figure~\ref{fig:CIICO}. Additional data from similar nearby spiral galaxies included in \citet{Hughes2017} and \citet{Gullberg2015} are also plotted as empty diamonds and squares, respectively.  We represent the regions in NGC~7331 with no CO emission or CO detections with SNR lower than 3 as upper limits (leftward arrows).  Over plotted is a grid of models showing the predicted $G_0$, the FUV radiation field in Habing units, and $n$, the gas density in cm$^{-3}$, values determined using PDR Toolkit \citep{Kaufman2006, Pound2008}.  The majority of the regions in NGC~7331 fall within the predictions of PDR Toolkit.  Most of the regions included in our sample show density conditions between 10$^4$ and 10$^5$~cm$^{-3}$, typical of PDRs, and $G_0$ of $\sim 10^2$, which is unsurprising for the relatively low star--formation rate observed in NGC~7331.

A few regions in NGC~7331 do not fall within the predictions made by the PDR Toolkit models (see green and cyan points in the un--shaded area of Figure~\ref{fig:CIICO}).  These regions represent areas where little to no CO emission was detected, indicating the possible presence of CO--dark gas.  This is highlighted by the fact that all of these points are upper limits. The majority of the points that fall well--beyond the main locus of the data from NGC~7331 also show low SNR values for the CO detections.  Interestingly, all of the regions included in the `Intra--Arm' (see Figure~\ref{fig:SEDfits}, right panel) had no clear CO detection, but detectable [CII] emission.  The lack of CO emission could be a sign of CO--dark molecular gas in this one region of NGC~7331, which suggests that using CO to trace molecular gas in galaxies could be environmentally dependant.

\section{Summary and Conclusions}
\label{sec:conclusions}
We have presented new FIFI-LS observations of the inclined spiral galaxy NGC~7331 at the [CII]~158~\micron\ line.  By comparing this new [CII] map with archival UV--IR measurements, we are able to trace a variety of ISM conditions across the nucleus, molecular ring, spiral arms, and disk of NGC~7331.  We find that:
\begin{enumerate}
    \item The fraction of the [CII] emission originating in the neutral ISM is higher when we view the external part of the molecular ring compared to when we view the internal part of the molecular ring.  This is likely due to the UV light from young stars in the nucleus ionizing the gas in the internal side of the ring, whereas dust within the ring blocks ionizing UV light from penetrating to the external side of the ring.
    \item There is a clear trend between the amount of attenuated UV emission and observed [CII] emission, with the molecular ring showing both the brightest [CII] emission and the largest amount of attenuated UV light.  The average [CII]/UV$_{\rm{Atten}}$ is 0.013, suggesting a photoelectric heating efficiency of 1.3\% throughout NGC~7331.
    \item The [CII]/FIR decreases with FIR surface density, extending the [CII] deficit trend observed in the KINGFISH sample of star--forming regions in local universe galaxies and the GOALS sample of local U/LIRGS to lower $\Sigma_{\rm{FIR}}$ values.  The clear continuation of the trend observed in the KINGFISH and GOALS sample to the local measurements across NGC~7331 highlight how the [CII] deficit effects measurements at a wide variety of scales.
    \item After correcting for analysis difference, the [CII]/PAH ratio measured across NGC~7331 matches the [CII]/PAH ratios measured in other local-universe galaxies.  The slight decreasing trend in [CII]/PAH with increasing $(\nu F_{\nu})_{70\mu m}/(\nu F_{\nu})_{100\mu m}$, an indicator of dust temperature, suggests differences in heating and cooling rates across the ring, spiral arms, nucleus, and disk of NGC~7331.  This could be caused by the increasing fraction of charged PAHs lowering the overall heating efficiency.
    \item Based on comparisons between the [CII] and CO emission, we find the majority of the neutral--ISM [CII] arises co--spatially with the CO gas.  Of interest, there are a few regions with [CII] detections without CO emission, especially along the area labelled `Intra--arm'.  This suggests the presence of CO--dark molecular gas.
\end{enumerate}
All of these results highlight the importance of understanding the environmental dependence of [CII]~158~\micron\ line measurements.  The comparisons between the \CII\ emission and the tracers of dust heating (FIR, PAH, and UV attenuation) show how the \CII\ emission from different environments behaves as an ISM coolant.  The lower values of \CII/FIR, \CII/PAH, and \CII/UV$_{\rm{Attenuation}}$ measured in the core compared the higher values measured in the disk suggest the local environment can have significant impacts on the photoelectric heating efficiency, potentially caused by the changes in grain charge observed in the PAH$_{7.7\mu m}$/PAH$_{11.3 \mu m}$ values across the disk.  The smooth transition between the global U/LIRG measurements of \CII/FIR to the local measurements from within NGC~7331, even to the most diffuse regions with low FIR surface brightnesses, shows how the \CII\ deficit is not just an effect of the extreme conditions in starburst galaxies.  The observations of \CII\ emission can also be influenced by observing perspective, as shown by the variation in the fraction of the observed \CII\ emission originating in the neutral ISM across the two sides of the molecular ring.  Taken together, the analysis of NGC~7331 presented in this work add to our understanding of the multitude of factors that can influence the strength of the \CII\ emission from other galaxies.  

As this emission line has become a frequent target of a multitude of high and low--z studies, it is essential we build a thorough understanding of how galactic conditions can impact the strength of [CII] emission.  By using the power of SOFIA to obtain spatially--resolved measurements of the [CII] emission from local--universe galaxies, we can better uncover how properties like inclination and the presences of rings and arms could effect the capability of this line to inform us about ISM conditions in distant galaxies.  We plan to expand this work by performing similar analysis on a set of nearby galaxies recently mapped by FIFI-LS onbard SOFIA.  By comparing the environments within NGC~7331 to similar environments in other galaxies, we can continue to build our understanding of the multitude of factors that can effect the strength of the [CII] line, and therefore potentially help or hinder future studies of [CII] emission from galaxies near and far.

\begin{acknowledgments}
We thank the anonymous referee for the helpful comments and constructive remarks which helped us improve the  presentation of our paper. This research is based on data and software from these projects: the NASA/DLR Stratospheric Observatory for Infrared Astronomy (SOFIA) jointly operated by USRA, under NASA contract NNA17BF53C, and DSI under DLR contract 50 OK 0901 to the University of Stuttgart; `The HERA CO-Line Extragalactic Survey’~\citep{Leroy2009} done with the IRAM~30m telescope; Herschel, an ESA space observatory science with instruments provided by European-led P.I. consortia and important NASA participation; the Spitzer Space Telescope, operated by JPL, Caltech under a contract with NASA; the SDSS survey, funded by the A. P. Sloan Foundation, the Participating Institutions, NSF, the U.S. Dep. of Energy, NASA, the Japanese Monbukagakusho, the Max Planck Society, and the Higher Education Funding Council for England. The Two Micron All Sky Survey (2MASS), a joint project of the University of Massachusetts and IPAC/Caltech, funded by NASA and NSF; GALEX, a NASA small explorer, whose archive is hosted by the High Energy Astrophysics Science Archive Research Center (HEASARC). Partial financial support for this project was provided by NASA through award \# SOF-06-0032 issued by USRA.
\end{acknowledgments}

%

\vspace{5mm}
\facilities{SOFIA (FIFI-LS), Herschel (PACS, SPIRE), Spitzer (MIPS, IRAC), ISO (ISOCAM), IRAM 30m, BIMA, VLA, GALEX, SDSS, 2MASS}


\software{astropy \citep{2013A&A...558A..33A,2018AJ....156..123A},  
sospex (\citet{2018AAS...23115011F}, \url{www.github.com/darioflute/sospex}),
        HIPE \url{www.cosmos.esa.int/web/herschel/hipe-download}, CIGALE (\citet{Boquien2019}, \url{https://cigale.lam.fr/})
          }


 
\appendix
\begin{figure*}[!b]
\begin{center}
\includegraphics[width=0.95\textwidth]{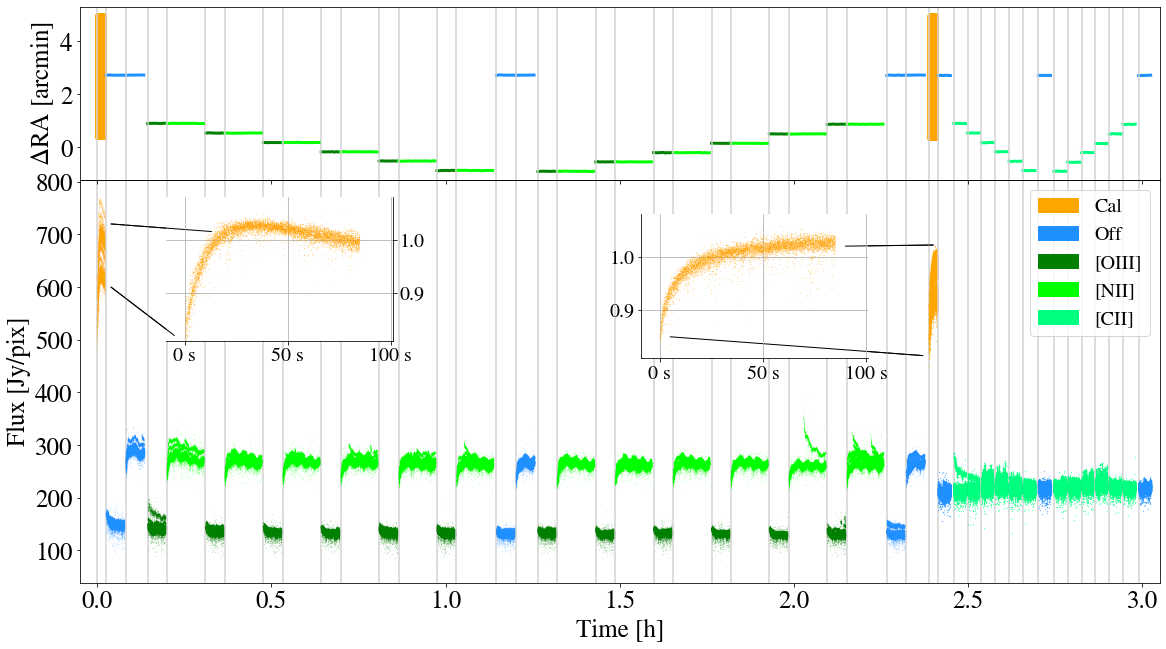}
\end{center}
\caption{Fluxes detected by the 16 spectral pixels of the PACS red array central module during the consecutive observations of the [NII]~122~$\mu$m, [OIII]~88~$\mu$m, and [CII]~157~$\mu$m.
The top panel displays the distance in R.A. of the array from the center of the galaxy to show how this raster observation was performed. The bottom panel is divided with vertical lines in observation blocks. Each block corresponds to a position of the array and a wavelength band.
Special blocks are the calibration blocks (orange) which consist in observing the two internal black bodies at the beginning of each AOR, and the blue blocks when the reference off position is observed. The green blocks correspond to the on-source observations. Blowouts of the calibration blocks, where the signal of each pixel has been normalized to its average value, show that transients affect the signal until the end of the block.
}
\label{fig:PACSobservation}
\end{figure*}
\section{PACS data reduction}
\label{sec:PACSreduction}

As mentioned in Section~\ref{sec:PACS}, the PACS data used in this paper have been taken in the so called {\it unchopped} mode. This mode is used when it is not possible to find an empty reference field close enough to the source to subtract from the on-source observation using the ``chop--nod'' method. In these cases, the reference field is observed at a distinct time and subsequently subtracted from the on-source observation. This method of observation can be affected by the known transient variation of the PACS response function.  These transient variations require more complex reductions since the response during the observation of the reference field could differ with respect to that during the observation of the source. In this section, we detail the reduction of the data used in the cases of the [NII]~122~$\mu$m and [CII] lines to show how they differ from the standard reduction which is used for the spectral cubes found in the Herschel Science Archive and explain why it is important to use such an advanced technique to obtain good calibrated fluxes from unchopped observations.

Figure~\ref{fig:PACSobservation} shows the two consecutive observations of the [NII]~122~$\mu$m, [OIII]~88~$\mu$m, and [CII]~157~$\mu$m lines with PACS. Fluxes from the 16 spectral pixels in the central module of the red channel of PACS are shown as a function of time. Each AOR starts with a calibration block, colored in orange in Figure~\ref{fig:PACSobservation}, when the two internal black bodies are alternately observed at key wavelengths close to the wavelength range observed in the AOR. In particular, the key wavelengths are 150~$\mu$m and 180~$\mu$m before the observations of the [NII]~122~$\mu$m and 
[CII]~157~$\mu$m lines, respectively.
These black bodies of known flux are used to calibrate the observed sky and source fluxes. In the insets of the bottom panel of Figure~\ref{fig:PACSobservation}, fluxes for each pixel of the calibration blocks are normalized to their mean values.  The transient in the response of the detectors due to sudden changes in flux is clearly visible in these insets where the telescope shifts from observing the background to the internal black--bodies. The calibration blocks are not long enough to reach the stabilization of the response. In particular, the first calibration block ends around the average value which is probably close to its asymptotic value. In the second case, the flux is monotonically increasing and it is already higher than the average at the end of the block. Since the standard calibration uses average values of the calibrators, the calibrator fluxes will be underestimated. To calibrate the observation fluxes are divided by this value and multiplied by the known black body fluxes. Therefore this leads to overestimated source fluxes.

Finally we note that, in order to avoid moving the telescope, the two wavelength bands in the first AOR (observing the [NII] and [OIII] lines) are observed sequentially for each raster position. The sudden flux changes between blocks introduce further transients in the response of the detectors.

\begin{figure*}[b!]
\begin{center}
\includegraphics[width=0.9\textwidth]{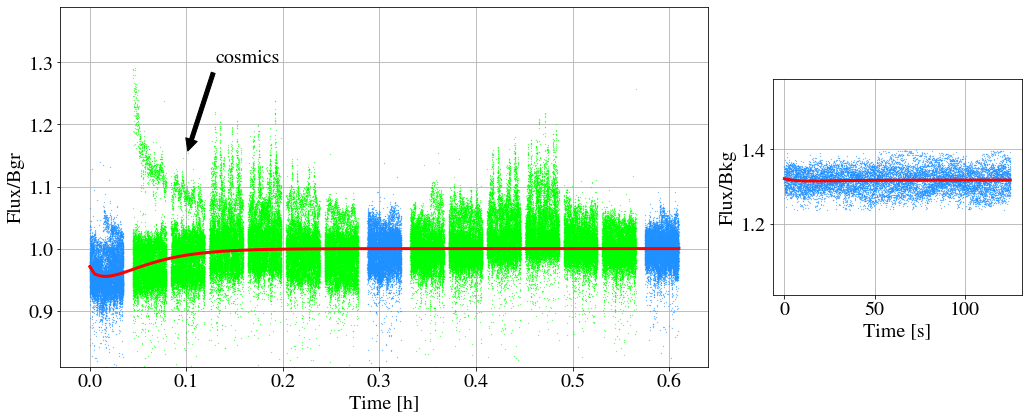}
\end{center}
\caption{Long term transient of pixels in the central spaxel of the [CII] line PACS observations. Each block with green points corresponds to a raster position across the observed strip, while the blue blocks correspond to the sky reference positions. The signal from each block has been divided by the telescope background and the level computed in last block (see right panel). In this case, the [CII] line calibration based on the calibration blocks is off by 30\%. The red line in the left panel is the fitted response. Cosmic ray add further transients to a few pixels which shows a marked different behaviour with respect to the average.
}
\label{fig:CIItransient}
\end{figure*}
\subsection{[CII] data and cross-correlation with FIFI-LS}
\label{sec:PACSvsFIFI}

Part of NGC~7331 has been observed with PACS at 158~$\mu$m.
This makes possible a direct comparison between fluxes detected with PACS on Herschel and FIFI-LS on SOFIA.

A way to study the variation of the response during an observation it to normalize the observed flux by the telescope background flux. This normalization was modeled and computed for each PACS observational day and it is routinely used to calibrate the `chop-nod' PACS observations.  In the reference off positions, one expects to observe only telescope background. For our observations, the continuum is negligible with respect to the telescope background. Therefore, the ratio of flux and telescope background should be close to one if the flux is well calibrated.

\begin{figure*}[b!]
\begin{center}
\includegraphics[width=0.9\textwidth]{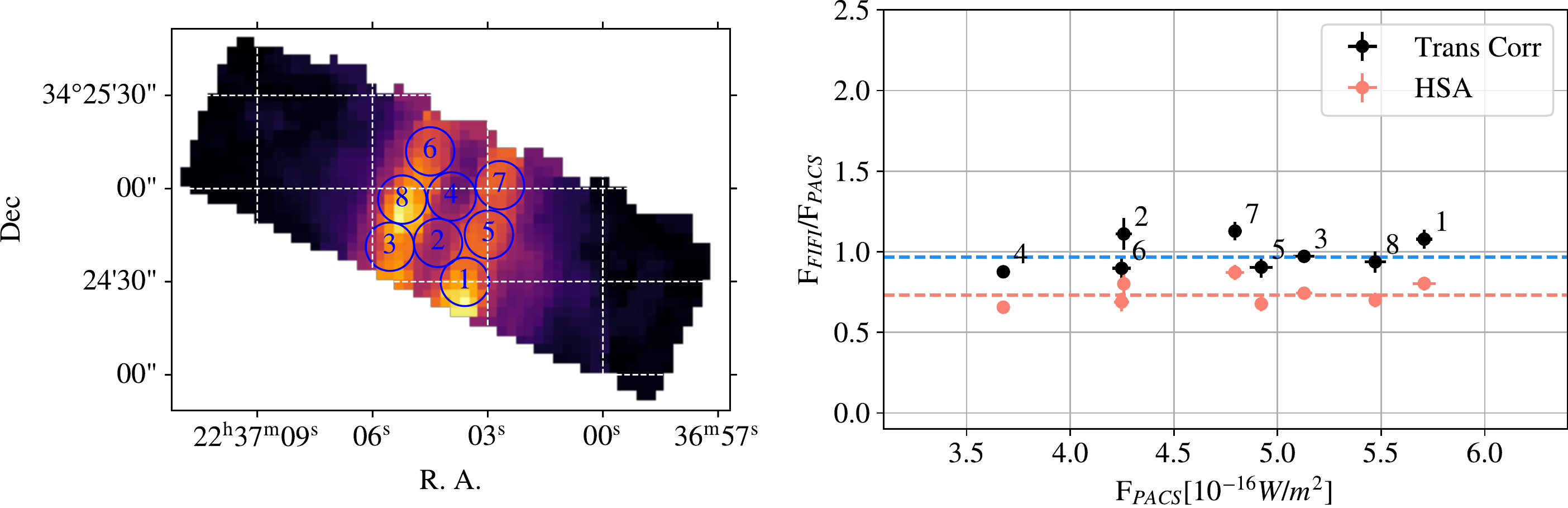}
\includegraphics[width=0.9\textwidth]{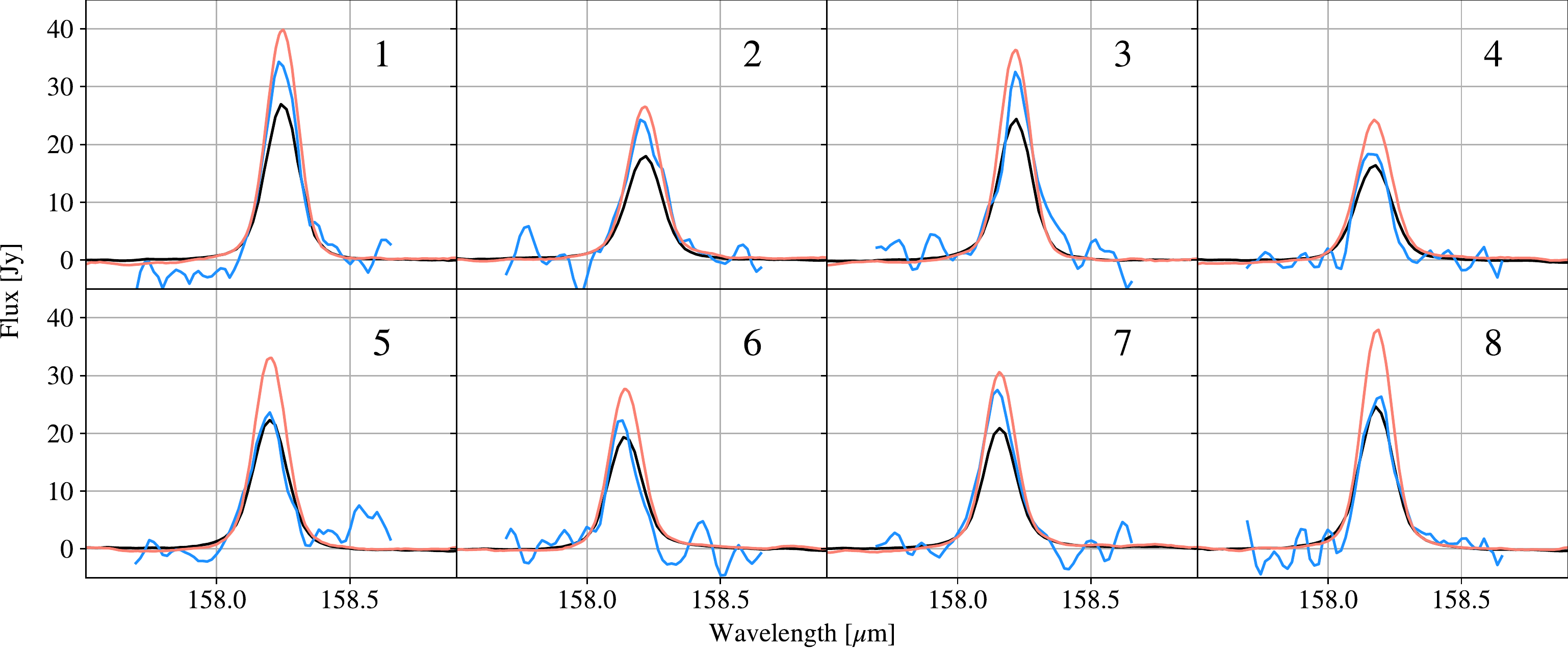}
\end{center}
\caption{
Comparison of fluxes measured with PACS and FIFI-LS on the common region.
In the top left figure, the eight chosen apertures are plotted over the integrated [CII] flux observed with PACS. The ratio of the fluxes in the same apertures is shown in the top right panel considering our reduction (Trans Corr) and the archival products (HSA). The mean of the ratios weighted with errors are marked with dashed lines.
In the bottom panel, the PACS and FIFI-LS spectra are after subtraction of the continuum. The FIFI-LS spectra are drawn in blue,
while the PACS spectra are shown in black (transient corrected) and salmon pink (HSA products).
The noisy nature of the FIFI-LS signal around the line adds some uncertainty to the flux measurement. We used the 160~$\mu$m image to guide our choice of continuum. Apertures are labelled with numbers from 1 to 8.
}
\label{fig:FIFIvsPACS}
\end{figure*}
We have already seen in Figure~\ref{fig:PACSobservation} that the sudden flux variation passing between calibration block and telescope background causes a transient variation of the response. Therefore, the best way to check the calibration against the background is to consider the flux measured during the last observation of the off reference during the AOR which should be the least affected by the transient. In the inset of   Figure~\ref{fig:CIItransient}, we can see that the data in the last off reference position are approximately 30\% higher than the telescope background. If we divide each block by this ratio, the long term transient clearly appears as the response gradually shifts over the total exposure time. These effects are taken into account by the transient correction pipeline in HIPE~15 which is used for our reduction. The products retrieved from the Herschel archive do not use the transient correction pipeline. As a consequence, they show a gradient along the strip which is due to the long term transient and, more importantly, have fluxes 30\% higher than the ones measured from transient corrected cubes calibrated to the telescope background.

When comparing PACS and FIFI-LS data, we have to consider strengths and problems of the two instruments. PACS unchopped mode does not allow to recover the absolute value of the continuum accurately. On the other hand, the baseline around the [CII] line in the FIFI-LS data is usually not wide enough to clearly define the continuum. The combination of lower exposure at the limits of the wavelength band and presence of telluric features makes the definition of the continuum challenging. For this reason, we based our fit on values inferred by broad band measurement (160~$\mu$m or SED fitting).

To compare the two measurements we degraded the spatial resolution of the PACS data to the one of the FIFI-LS data. Then, we defined a series of eight apertures, the size of the FIFI-LS beam (diameter of 15.6~arcsec), along the ring  and inside it (see Fig.~\ref{fig:FIFIvsPACS}). After subtracting the continuum, we fitted the lines with pseudo-Voigt profiles. Looking at the PACS spectra (black and salmon lines in the bottom panel of Figure~\ref{fig:FIFIvsPACS}), it is clear that the wings of the line extend all over the FIFI-LS coverage. To have a fair comparison, we measured the line flux within a distance of one FWHM from the line center. The comparison clearly shows that the fluxes from the HSA products are too high in comparison with fluxes measured with FIFI-LS, while the spectral cube obtained with the transient correction pipeline compares well with the FIFI-LS cube. The average of the flux ratios between our reduction of the PACS data and the FIFI-LS data is $0.97 \pm 0.02$, taking into account the errors on the fluxes. When considering the data available in the Herschel Science Archive which were not corrected for transients and calibrated using the calibration block only, the ratio is $0.73 \pm 0.01$. 
Since the flux calibration of FIFI-LS has been successfully validated by cross-correlations with PACS observations taken in `chop-nod' mode, we conclude that if data are not calibrated using the telescope background as absolute reference and corrected for transients, this can lead to errors up to 30\% in the flux of PACS sources.

\begin{figure*}[b!]
\begin{center}
\includegraphics[width=0.95\textwidth]{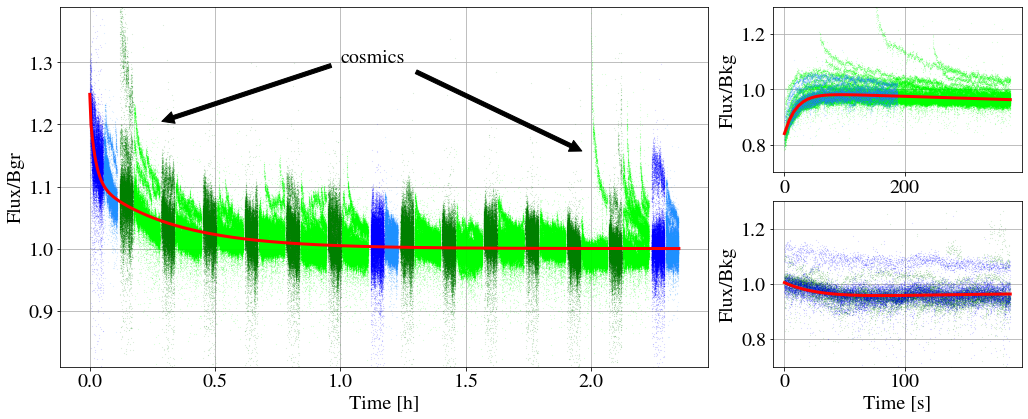}
\end{center}
\caption{Long term transient through the observations of all the pixels of the central spaxel of the [NII] line PACS observations of NGC~7331. Blocks are color coded as in Figure~\ref{fig:PACSobservation}. The two different tonalities of green and blue refer to the two lines observed in the observation: light for the [NII]122~$\mu$m and dark for the [OIII]88~$\mu$m red parallel data. The signal from each block has been divided by the telescope background and the band--switching induced transient computed using the last observation blocks (see right panels). The red line is the fit to the long--term transient which is applied to correct the flux level of the observation. Further transients, caused by cosmic ray hits, affect a few pixels (see points indicated with cosmics). Such transients are identified and corrected by the pipeline.
}
\label{fig:NIItransient}
\end{figure*}
\subsection{[NII]~122~$\mu$m data}
\label{sec:NIIdata}

The reduction of the PACS observations of [NII]~122~$\mu$m line required particular care. In fact, they are not only taken in unchopped mode, but also chained with observations of another line ([OIII]~88~$\mu$m).
As visible in Figure~\ref{fig:PACSobservation}, which displays the fluxes detected by the 16 pixels in the central module of the PACS array, the observation starts with a calibration block, followed by an observation of a reference off position, and then by pointings along the galaxy. The cycle is repeated twice ending with another observation of the reference off position. For each position of the telescope, the two wavelength bands are observed sequentially in order to reduce the time lost by slewing the telescope.
This method of switching between wavelength bands has the unfortunate consequence of introducing transients in each observational block. In the right panels of Figure~\ref{fig:NIItransient} we can see the last blocks observed normalized to the telescope background. Since the continuum of the galaxy in the external parts is very weak, this block sees practically only the telescope background and therefore can be used (along with the last reference off) to obtain a robust measure of the transient response along the block.  Using the available model of the telescope background available through HIPE to normalize the data, the effects of the transient variations in these bands is clearly visible. It appears clear that each [NII] observation (shown in bright green) is affected by a lower response at the beginning which gradually stabilizes during the block. We also notice that the length of the reference off (shown in blue) is half that of the on-source observation. This means that the reference off is even more affected by the transient than the on-source observation as the time required for the stabilization of the response function is barely reached.
The other wavelength band, parallel to the [OIII]~88~$\mu$m observed in the blue and shown as dark green points in Figures~\ref{fig:PACSobservation} and \ref{fig:NIItransient}, is affected by transients in the opposite way. The response is slightly higher at the beginning and stabilizes during the observation. 

\begin{figure*}[b!]
\begin{center}
\includegraphics[width=0.8\textwidth]{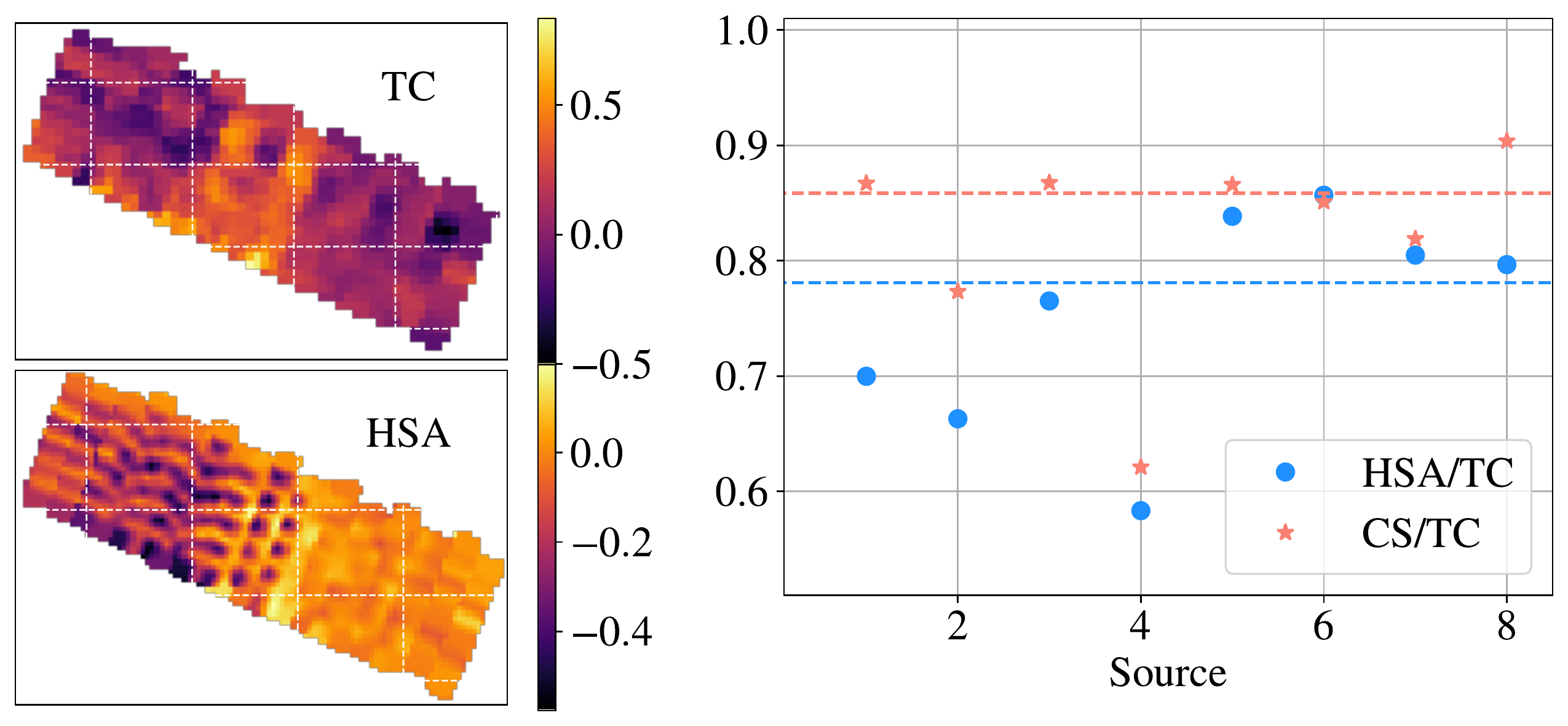}
\includegraphics[width=0.8\textwidth]{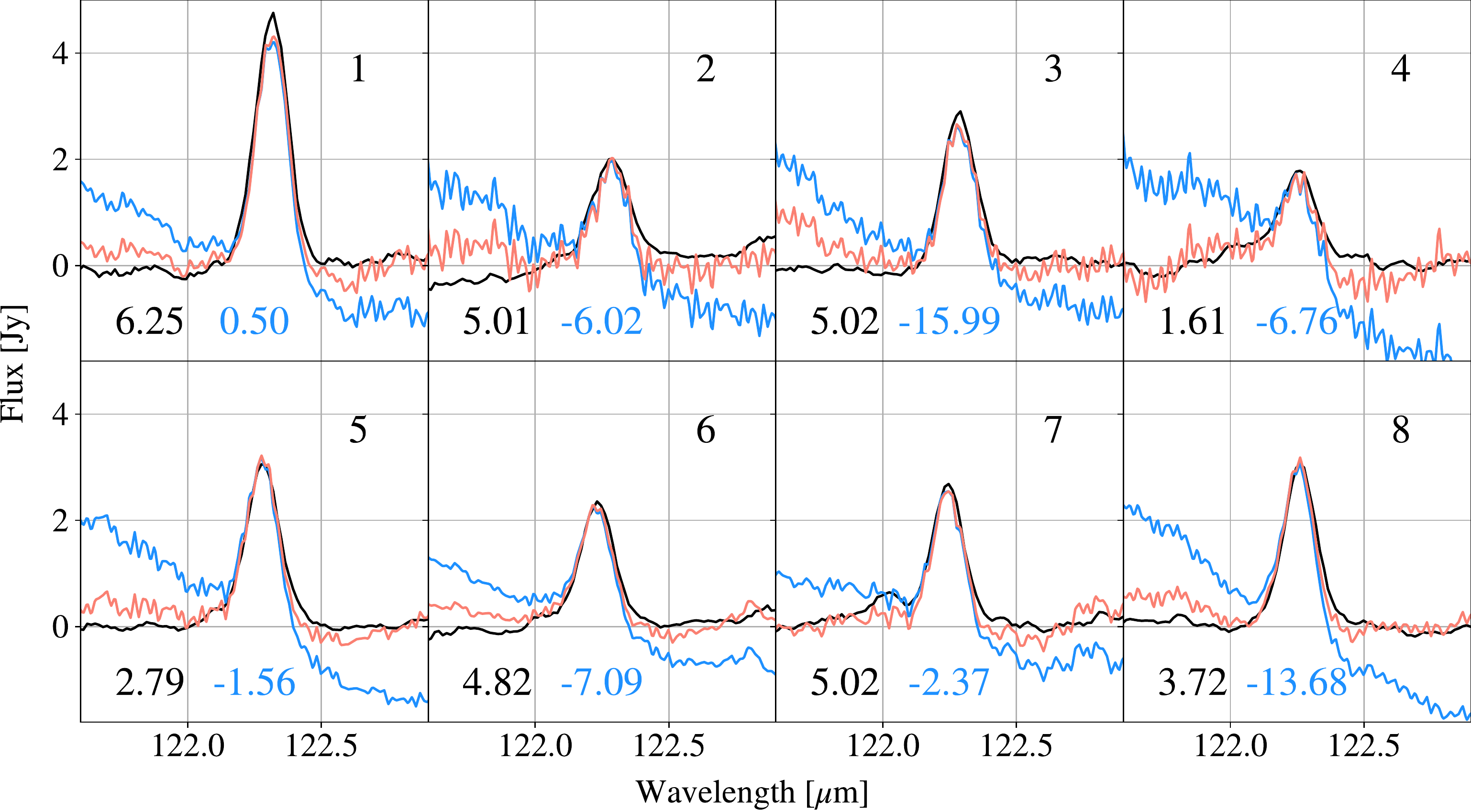}
\end{center}
\caption{
Comparisons of [NII]~122~$\mu$m fluxes measured in cubes from the Herschel archive (HSA) and reduced with the transient correction (TC) pipeline. In the top left panel, the slice of the cubes corresponding to the wavelength of the [NII] line is shown (flux is in Jy/pixel). Significant gradients are visible in the HSA image, while TC image is flatter and brighter on the ring location. The ratios of fluxes in the same apertures as in Figure~\ref{fig:FIFIvsPACS} are displayed in the top right panel. The ratios are also shown for the case when the continuum is subtracted from each spatial pixel before extracting the aperture (CS). Medians of the ratios are marked with dashed lines. In the bottom panel, spectra after subtraction of the continuum under the line are shown (HSA, CS, and TC in blue, black, and pink, respectively). The HSA products are much noisier than the TC spectra. Values of the continua under the line are reported under each spectrum.
}
\label{fig:NIIcomparison}
\end{figure*}
In addition to the transients introduced by the band switching, the PACS observations are also affected by long--term transient behavior. The presence of the calibration block when the PACS arrays sees the internal black bodies is another source of transients in the observations. As the flux seen by the PACS array during the calibration is more than double the flux during the observation, the jump in continuum between the calibration and the observations causes a significant transient effect. This, and occasionally the memory of the previous observation, affect the response of the array for most of the observation. As visible in Figure~\ref{fig:NIItransient}, a long-term transient can be fit during the observation to correct for the variation of flux level due to the variation of sensitivity of the array. After correcting each block for the variation of the response due to the band change, this long term transient becomes very clear. The variation of flux during this particular observation is around 20\%.
Since different positions along the plane of the galaxy have been observed at different times, not correcting for such a transient response introduces a gradient in the direction of the scan which is repeated twice in opposite directions. Also, since the reference off position is subtracted from the on-source position, the spectral cube will have negative offsets if the off position is not corrected.

By dividing the signal by the variable response of the detector during the observation, the fluxes are corrected during the entire observation.  As shown in Section~\ref{sec:PACSvsFIFI}, this correction of the response is important to obtain correct fluxes for unchopped PACS data. 

The transient correction pipeline also considers transients induced by cosmic rays hitting the detectors (see for instance the outliers in Figure~\ref{fig:NIItransient}). Such transients are detected and corrected. We stress that the data products available in the Herschel archive are not corrected for any transient effects.

Contrary to the case of [CII], the calibration of the [NII]~122~$\mu$m observation with the telescope background does not significantly differ from that using the calibration blocks. However, the change in background caused by transients can still bias flux measurements. Figure~\ref{fig:NIIcomparison} shows the comparison between fluxes measured in the same apertures along and inside the ring of the galaxy using spectral cubes from the Herschel archive (HSA) and the one reduced by us with the transient correction pipeline (TC). If we simply extract apertures in the HSA cube, we will find fluxes 78\% smaller than those measured in the TC cube. Most of this is due to the coaddition of pixels with different continua. Since the raster scan along the galaxy strip was repeated twice, the final spectral cube is obtained by combining spectra with different underlying continua. The slices of the spectral cubes in Figure~\ref{fig:NIItransient} show the dramatic difference between the HSA and TC cubes. The HSA cube has a large gradient across the scan direction of the raster and ripples along the perpendicular direction. The TC cube has a flatter background and shows some continuum flux along the ring. The spectra extracted in the apertures have steep slopes in the continuum and a noise higher than that of the spectra extracted from the TC cube. Since the continuum of HSA cube varies significantly among the different spatial pixels, we can mitigate the effect of coadding spectra with different continua in the apertures by subtracting from each spatial pixel the continuum fitted around the line before the extraction. The spectra extracted from this continuum subtracted (CS) cube result in a better agreement with the fluxes from the TC cube: the ratio is 86\%. Nevertheless, the quality of the HSA spectral cube is already compromised by the coaddition of spectra with different continua taken during the two scans of the region. It is therefore paramount to perform a careful correction of transients in the unchopped PACS mode to avoid biases in the flux measurements.

\bibliography{main}{}
\bibliographystyle{aasjournal}



\end{document}